\documentclass [11pt, letterpaper]{article}
\pdfoutput=1 
\usepackage{fullpage}
\usepackage{amsmath}
\usepackage{amssymb}
\usepackage{graphicx}
\usepackage{braket}
\usepackage{mathrsfs}
\usepackage{wrapfig}
\usepackage{boxedminipage}
\usepackage{epsfig}
\usepackage{setspace}
\usepackage{subfigure}
\usepackage{float}
\usepackage{hyperref}
\usepackage{enumerate}
\usepackage{cite}
\usepackage{bbold}
\usepackage[font=footnotesize,labelfont=rm]{caption}

\def\Tr{{\rm Tr}}

\begin{document}

\begin{flushright}
{\tt arXiv:1905.$\_\,\_\,\_\,\_\,\_$}
\end{flushright}

{\flushleft\vskip-1.35cm\vbox{\includegraphics[width=1.25in]{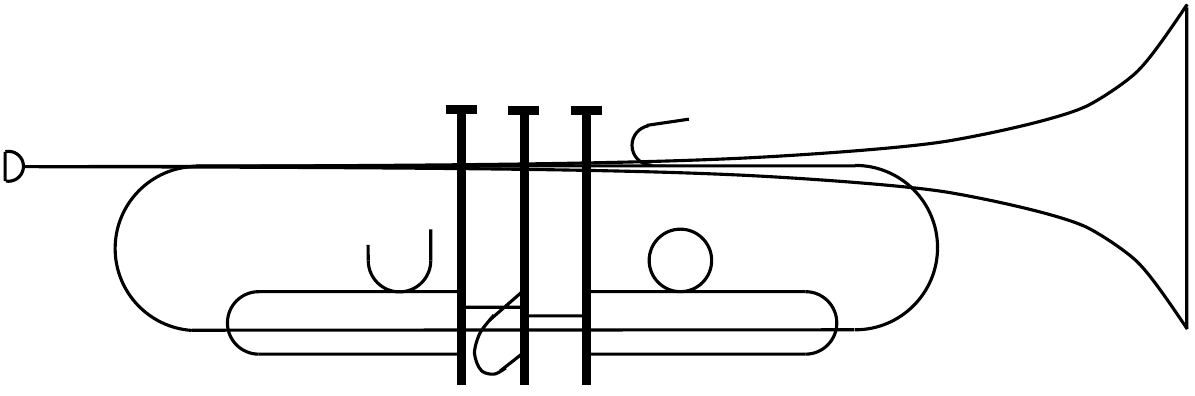}}}

\bigskip
\bigskip

\bigskip
\bigskip
\bigskip
\bigskip

\begin{center} 

{\Large\bf   Holographic Heat Engines as Quantum Heat Engines}

\end{center}

\bigskip \bigskip \bigskip \bigskip

\centerline{\bf Clifford V. Johnson}

\bigskip
\bigskip
\bigskip

\centerline{\it Department of Physics and Astronomy }
\centerline{\it University of
Southern California}
\centerline{\it Los Angeles, CA 90089-0484, U.S.A.}

\bigskip

\centerline{\small {\tt johnson1}  [at] usc [dot] edu}

\bigskip
\bigskip


\begin{abstract} 
\noindent 
Certain solutions of Einstein's equations in 
anti--de Sitter spacetime can  be engineered, using extended gravitational thermodynamics, to yield ``holographic heat engines'', devices that turn heat into useful mechanical work. 
On the other hand, there  are  constructions (both experimental and theoretical) where a series of operations is performed on a small quantum system, defining what are known  as  ``quantum heat engines". We propose that certain holographic heat engines can be considered models of quantum heat engines, and the possible fruitfulness of  this connection is discussed. Motivated by features  of quantum heat engines that take  a quantum 
system through  analogues of certain classic thermodynamic cycles,  some black hole Otto  and Diesel cycles are presented and explored for the first time. In the expected regime of overlap, our  Otto efficiency formulae are of the  form exhibited by  quantum  and classical heat engines. 
\end{abstract}

\pagenumbering{gobble}

\newpage 

\pagenumbering{arabic}

\baselineskip=18pt 
\setcounter{footnote}{0}

\section{Introduction}
\label{sec:introduction}

\subsection{Background}
\label{sec:background}

The thermodynamics of heat engines, refrigerators, and heat pumps is often thought  to be firmly the domain of large classical systems, or put more carefully, systems that have a very large number of degrees of freedom such that thermal effects dominate over quantum effects. Nevertheless, there is thriving field devoted to the  study---both experimental and theoretical---of the thermodynamics of machines that use small quantum systems as the working substance. (We shall say {\it  heat engines} as a shorthand henceforth, but everything we say here can apply to  heat pumps and refrigerators.) 

Connecting the framework of heat engines to intrinsically quantum systems goes back as far as  1959's 
ref.~\cite{PhysRevLett.2.262}, where  the  three--level maser was reimagined as a continuous heat engine. Both continuous and  reciprocating heat engines where  the working substances are undeniably  quantum in operation are now widely studied, and  have relevance to a variety of  subjects of practical concern, such as the physics and engineering of quantum devices, the design and control of qubits for use in quantum information, and the concomitant broad field of open quantum systems\footnote{See refs.\cite{breuer2002theory,doi:10.1080/00107514.2016.1201896,Halpern:2018bul} for excellent recent reviews.}. 

Viewed as a cycle in thermodynamic state space, the reciprocating heat engine (whether quantum or classical)  is a closed loop of processes performed on the  central apparatus that involves coupling to  external systems. These systems  bring about state changes that may or may not involve heat exchange. The heat exchanges result from a coupling to hot and cold heat baths (at temperatures $T_H$ and $T_C$ respectively, so that heat $Q_H$ (resp. $Q_C$) flows in (out).  The resulting work done follows from the first law of thermodynamics: $W=Q_H-Q_C$. The primary figure of merit  characterizing this engine is the efficiency $\eta=W/Q_H=1-Q_C/Q_H$. The second law of thermodynamics ensures that it is bounded above by the Carnot efficiency $\eta_{\rm C}^{\phantom{C}}=1-T_C/T_H$.

In a seemingly unrelated corner of physics, a black hole in asymptotically anti--de Sitter spacetime, when quantum effects are taken into account, acts as a thermodynamic system\cite{Bekenstein:1973ur,Bekenstein:1974ax,Hawking:1974sw,Hawking:1976de,Hawking:1982dh} with internal energy $U$ (set by the   black hole mass $M$), temperature $T$ (the surface gravity divided by~$2\pi$), and entropy~$S$ (1/4 of the horizon area).\footnote{There are appropriate factors of the speed of light, $c$ and $\hbar, G,$ and $k_{\rm B}$, the constants of Planck, Newton, and Boltzmann, respectively} The first law is $dU=TdS$. This physics describes the high temperature sector of a (non--gravitational) quantum system, the (generalized) gauge theory to which the gravitational physics  in AdS is ``holographically''\cite{tHooft:1993dmi} dual\cite{Maldacena:1997re,Witten:1998qj,Gubser:1998bc}. For example, in $4{+}1$ spacetime dimensions,  the dual theory is an $SU(N)$ gauge theory in 3{+}1 dimensions. Crucially, when reliable gravity computations can be done ({\it i.e.,} curvatures are low),~$N$ is large. Since $N^2$ measures the number of degrees of freedom in the theory, we see that this is the thermodynamics of a large classical system, despite the underlying theory being a quantum field theory.

 It is possible to embed all of this physics into a larger thermodynamics framework sometimes called  ``extended black hole thermodynamics''. The cosmological constant $\Lambda$ is treated as a dynamical variable\cite{Kastor:2009wy,Wang:2006eb,Sekiwa:2006qj,LarranagaRubio:2007ut,Henneaux:1984ji,Henneaux:1989zc,Teitelboim:1985dp} and defines a pressure $p\,{=}{-}\Lambda/8\pi G$, positive for asymptotically AdS spacetimes. 
 While $T$ and~$S$ have the same identification in terms of black hole properties, the mass~$M$ is now the enthalpy\cite{Kastor:2009wy} $H{=}U{+}pV$. The conjugate variable to $p$, the thermodynamic volume, emerges as $V{=}(\partial H/\partial p)|_S$. The first law is now the more familiar--looking $dU{=}TdS{-}pdV$.
 
 The presence of a mechanical work term $dW{=}-pdV$ allows for the definition of a traditional  reciprocating heat engine using the black hole as the working substance\cite{Johnson:2014yja}. Such a ``holographic heat engine'' is a closed cycle in thermodynamic state space, as before, with coupling to a hot heat reservoir at temperature $T_H$ and a cold heat reservoir at temperature $T_C$,  net heat flow $Q_H$ in ($Q_C$ out, respectively).  Again, its efficiency 
  is bounded from above by that of the Carnot engine.
 
Interpretations of such an engine   can be made in  the underlying (dual) quantum system which supplies the quantum description of the black hole microstates. Every point in state space is identifiable as a holographically dual non--gravitational quantum system\cite{Johnson:2014yja}. Moving from one point to another can be described in terms of processes entirely describable in terms of quantum field theory, such as thermalization and heat transfer (by coupling to a heat bath) or changing of couplings (perhaps by coupling to external systems or fields). 

\subsection{Synthesis}
\label{sec:synthesis}

It is easy to anticipate the next step in the chain of logic, given the ingredients above. Can one declare that holographic heat engines are quantum heat engines? In order to do that, we need one more crucial ingredient: The quantum system in play, the one coupled to the reservoirs, must in some sense be {\it small}. However, the quantum system in this context, the quantum field theory, is {\it large}, as we saw. It was a necessary condition for the duality to a black hole to make computational sense. We could try to make $N$ small, but that would require the gauge theory coupling to be large, in which case we lose computational control on that side.  So it would seem therefore that we cannot, while retaining sight of either theory, make the connection between holographic heat engines and quantum heat engines\footnote{It is worth remarking here that using a broader use of the term, the holographic heat engines presented in ref.\cite{Johnson:2018amj} are quantum heat engines in that they are directly built from processes that translate to deformations of entanglement in a quantum field theory. However, this it not what is usually meant in the literature, or in this paper.}. However, a recent observation adds a new ingredient. 

The point is that isochoric processes, for the correct choice of  black hole, have a tunably small window of energy states in play, as signalled by a Schottky--like peak in $C_V(T)$ discovered in ref.~\cite{Johnson:2019vqf}. {\it This is the ``small'' system that we seek.} It is a special subsector of the black hole degrees of freedom by can be naturally isolated in the extended thermodynamics. We will build  Otto engines\footnote{These are the first black hole Otto engines in the literature. Moreover, this means that all holographic heat engines presented thus far, while interesting, are not of the right type to compare well to quantum heat engines, since there are large numbers of degrees of freedom in play throughout the cycle.}, where the only heat exchanges are on isochores and therefore only these restricted states can be excited. These are, we propose, the black hole engines that are to be compared to the  Otto engines commonly used in the quantum heat engine literature.

Since we can tune that energy window/subsector to be as small as we like, {\it we propose that these holographic heat engines are models of quantum heat engines}\footnote{It goes without saying that the analogous refrigerators and heat pumps obtained by reversal of the cycle are defined by using the same ingredients.}. This proposed connection between these two disparate fields has a great deal of potential. A core pragmatic aspect of all this is the observation  that the gravitational physics ({\it e.g.,} that of a black hole), can supply  a rich variety of equations of state ---sometimes in closed form--- that can be engineered into a model of a quantum heat engine. This,  at the very least, gives an interesting and  powerful theoretical arena for exploring  models of interesting behaviour seen in quantum heat engines being studied in the laboratory. 

It is important to entertain the possibility of running the engine in a regime where quantum effects can make their presence known.  
Assuming a coupling to ordinary thermal reservoirs, as we do (it seems that this is  built into the  description on the black hole side)  there is no reason to expect the efficiency $\eta$ to not be bound by Carnot. However, it might be possible to engineer schemes by which the efficiency can be enhanced by special (possibly quantum) features of the working substance. Moreover, by considering the power as well as the efficiency (where the time taken to do a cycle might be intrinsically connected to quantum effects) we might learn new schemes for how quantum effects influence an engine's features. This is of great experimental and theoretical interest in the quantum heat engine field. Having  dual gauge theory descriptions of some of the engine's cycles might mean that holographic heat engines could well be a useful additional tool in this arena. We will not explore this aspect in this paper, but it is an important subject for future exploration.

In this paper we will  present and explore essential  aspects of our engines. The quantum heat engine framework (reviewed briefly in section~\ref{sec:quantum-heat-engines}) suggests what kind of black hole we ought to study in order to get access to an engine with somewhat analogous features. We will focus on  Otto cycles initially. The key property needed is a non--vanishing specific heat at constant volume,~$C_V$, allowing the system to have  distinct adiabats and isochoric processes. In section~\ref{sec:black-holes} we choose some black holes with this property and discuss some key useful features of the behaviour of their $C_V(T)$. We build   Otto engines in section~\ref{sec:black-hole-otto}, and discuss some of their properties. In section~\ref{sec:critical-engines} we present the results of a study of a suggestion in the quantum heat engine literature to run quantum heat engines near a critical point\cite{2016NatCo...711895C}, in order to see the effects on the efficiency (revisiting a study we did with a Brayton--like cycle in ref.\cite{Johnson:2017hxu}). A non--vanishing $C_V$ means that the Brayton--like cycles commonly used for holographic heat engines in the literature (starting with the prototype  of ref.~\cite{Johnson:2014yja}) can be made much richer, and so for completeness  we briefly discuss  genuine Brayton engines in section~\ref{sec:brayton-diesel}, along with Diesel engines\footnote{The exact efficiency formula for a Brayton black hole engine was presented in ref.\cite{Johnson:2016pfa}, and used as a building block for more general $C_V\neq0$ engines in ref.\cite{Chakraborty:2017weq}. See also ref.\cite{Hennigar:2017apu} for some discussion of engines made from $C_V\neq0$ black holes. Kerr black holes are used in those papers as illustrative examples, but they do not construct Otto and Diesel cycles, as we do here.}.   We end with a short discussion in section~\ref{sec:discussion}.

\section{Quantum Heat Engines}
\label{sec:quantum-heat-engines}

\begin{wrapfigure}{r}{0.45\textwidth}
\centering
\includegraphics[width=0.45\textwidth]{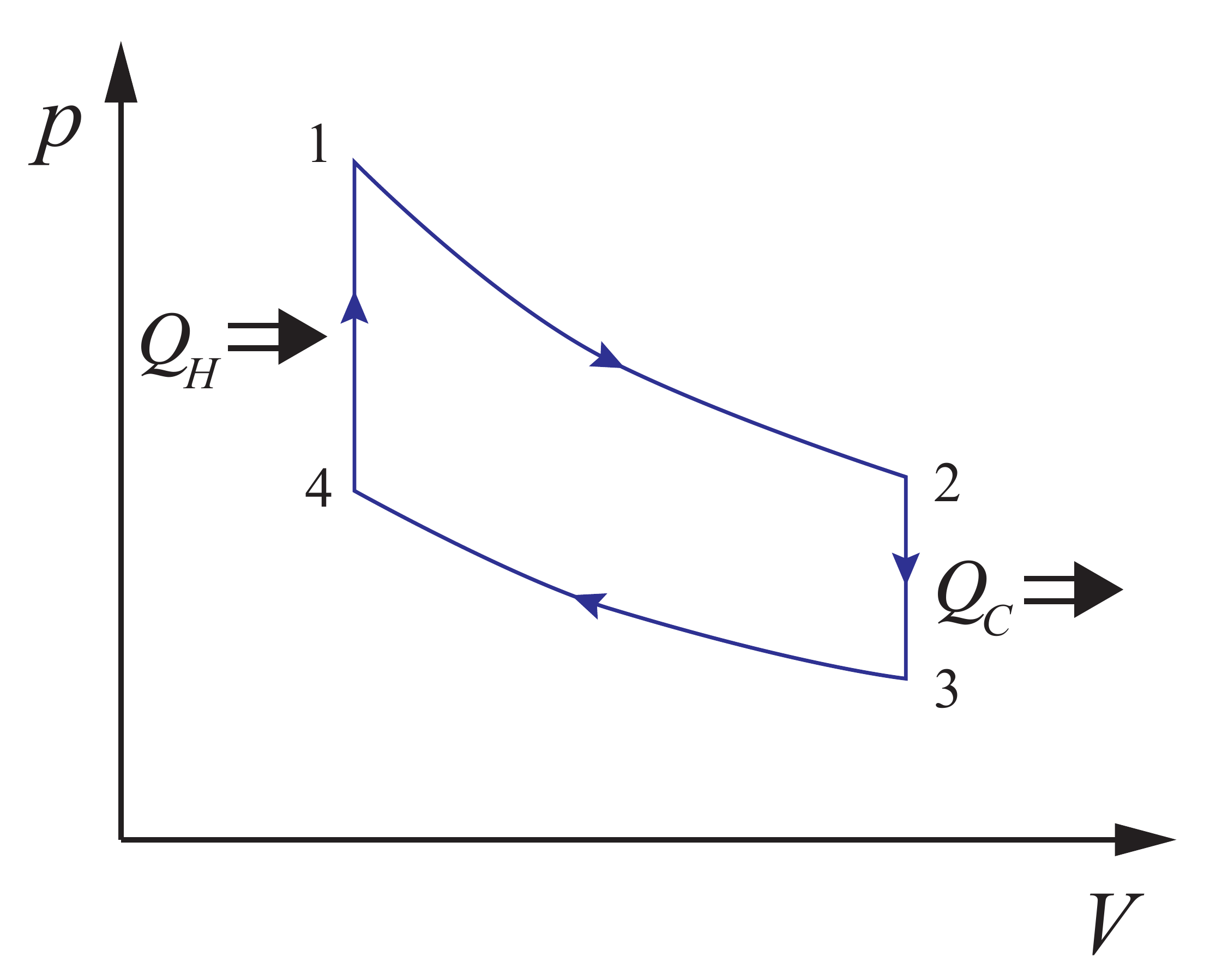}
\caption{\label{fig:Otto-cycle} The Otto engine cycle, made of two adiabatic strokes and two isochoric heat transfers.}
\end{wrapfigure}

A commonly discussed  quantum heat engine involves taking a simple quantum system (either discrete or continuous) through four steps, the fourth returning the system to its original state. Two of the steps are adiabatic, while the other two are isochoric, and are where the heat exchanges take place. This Otto engine cycle is shown in  the $p{-}V$ plane in figure~\ref{fig:Otto-cycle}, with labels for reference.
%
\begin{itemize}
\item Process $1 \rightarrow 2$ is an adiabatic expansion, the power stroke, where the engine  does work while dropping in temperature.
\item Process $2 \rightarrow 3$ is an isochoric heat exchange, where a quantity $Q_C$ of heat leaves the system, resulting in a further temperature drop.
\item Process $3 \rightarrow 4$ is an adiabatic compression, resulting in a temperature increase.
\item Process $4 \rightarrow 1$ is an isochoric heat exchange, where a quantity $Q_H$ of heat enters the system.
\end{itemize}
How the engine is realized depends upon the precise system, and the control parameters available. Overall, however, there's a quantum thermodynamics description of the energy of the system (see {\it e.g.} refs.\cite{Kieu2006,PhysRevE.76.031105,doi:10.1080/00107514.2016.1201896}) in terms of~$\rho$ and  ${\cal H}$, the state and  Hamiltonian: $U{=}\Tr[\rho{\cal H}]$. Work on or by the system corresponds to changing ${\cal H}$.  For example, for a harmonic oscillator it would amount to adjusting the basic frequency $\omega$ governing the spacing of the energy levels $E_n{=}(n+\frac12)\hbar\omega(t)$. Such a coupling is tunable in the laboratory (it might be set by the physical  geometry of a resonator, or by the fields in an  ion trap, or by the inductance or capacitance of a conducting microcircuit,  {\it etc}).  Heat flow to or from the system comes from the $\rho$ changing in response to the  coupling to the environment. Formally, on average these are:
\begin{equation}
\langle W\rangle=\int_{t_i}^{t_f} \Tr[\rho^{(t)}{\dot{\cal H}}^{(t)}] dt\ ,\quad {\rm and}\quad  \langle Q\rangle=\int_{t_i}^{t_f} \Tr[{\dot \rho}^{(t)}{\cal H}^{(t)}] dt\ ,
\end{equation}
In this language the meaning of an adiabatic change of the parameter (performed in steps 1-2 and~4-1) is that the precise pattern of energy occupation is unchanged, which is achievable for slow enough (isolated) evolution.  That the  changing parameter maps directly to volume changes follows from the first law, using that we're on an  adiabat: $dS{=}0$.  For the harmonic oscillator, changes in $\omega$ map to changes in volume, with $V\sim \omega^{-1}$. Expansion (compression) corresponds to a reduction (increase) in the basic oscillator frequency $\omega$ between the values~$\omega_h$ and~$\omega_c$, where $\omega_h>\omega_c$. This identification of a volume change  with a change in a basic coupling in the quantum system is an important clue for how to interpret the effective thermodynamic volume in gravitational systems, as we'll discuss later. 

The precise details of the heat exchange are also system dependent. In all cases they amount to putting the system into thermal contact with a larger system that can act as a sink or source of heat. An extremely common way to model this is to write ${\dot \rho}^{(t)}$ in terms of a coupling of $\rho$ to operators of Lindblad type\cite{breuer2002theory}, representing the environment, but this is not the only approach (see refs.~\cite{doi:10.1080/00107514.2016.1201896} for a review.).

For the quantum harmonic oscillator, if  the cycles can be performed extremely slowly, so as to achieve perfect adiabaticity and frictionless thermalization, all energy changes can be written in terms of overall frequency changes or population changes. Then the total work done amounts to\cite{Rezek_2006} $W=\hbar \Delta\omega \Delta N$, where $\Delta\omega = \omega_h-\omega_c$ and $\Delta N = \Delta N_h=\Delta N_c$ is the population change along the isochores. This is the difference in heat flow,  $W=Q_H-Q_C$, and so the efficiency is simply:
\begin{equation}
\label{eq:efficiency-harmonic-oscillator}
\eta=1-\frac{\omega_c}{\omega_h}=1-\frac{V_1}{V_2}\ .
\end{equation}
 If $\omega_c/\omega_h$ is chosen as $T_C/T_H$ then $\eta=\eta_{\rm C}^{\phantom{C}}$, the Carnot efficiency, but otherwise it is smaller. Either way, this is the ideal situation. Much of the study of quantum heat engines is concerned with modelling the non--ideal situation, especially the coupling mechanisms by which thermalization (perhaps with friction effects) takes place, working at finite power, and so on.

\section{Black Holes}
\label{sec:black-holes}
A key lesson from the previous section is that we need a system with an equation of state that allows for adiabatic processes that are independent of isochoric processes. This might seem like an obvious statement, but it is the central question here, since the simplest black holes (static, pure geometry) such as Schwarzschild and Reissner--Nordstr\"om solutions, do {\it not} have this independence. The entropy, $S$, given by a quarter of the area of the horizon, is a simple power of the horizon radius (denoted $r_+$ in what follows). The thermodynamic volume $V$ is given by the geometric (naive) volume of the space occupied by the black hole, which is also a power of $r_+$. Therefore isochores and adiabats are the same\cite{Dolan:2010ha,Johnson:2014yja}. In other words, we cannot use these black holes to make Otto engines. 

The above observation  is equivalent to saying that those black holes has the {\it constant volume} specific heat $C_V=0$. Since in a sense, $C_V$ is a  direct measure of the available degrees of freedom (at least 
the traditional  ones  that can be excited without making volume changes), we see that  Reissner--Nordstr\"om, Schwarzschild, (and  various other simple static black holes), do not have any usable degrees of freedom for our purposes. 

The answer lies in studying black holes whose thermodynamic volume $V$ is independent of the entropy $S$. These are slightly more complicated, but readily available. The simplest one is probably the Kerr--AdS black hole, obtained by adding a rotation parameter\footnote{Potentially, the charged BTZ black hole\cite{Banados:1992wn,Banados:1992gq,Martinez:1999qi} is even simpler, and the extended thermodynamics has been worked out for it in refs.\cite{Gunasekaran:2012dq,Frassino:2015oca}. In fact $C_V(T)$ can be worked out analytically, revealing (see an update of ref.\cite{Johnson:2019vqf}) that it is unfortunately negative for all $T$, presumably signaling some sickness in the physics. We thank Felipe Rosso for suggesting that we look at this example.}. The appearance of the resulting conserved angular momentum, $J$ in the thermodynamic parameters ensures that $C_V\neq0$. Another choice is the family of ``STU" black holes in AdS. These are charged black holes that are  coupled to four $U(1)$ gauge fields (and three scalars) in an asymmetric way. The familiar Reissner--Nordstr\"om black hole is a special case where the charges $Q_i$ ($i=1,\cdots4$) are all equal  (and the scalars decouple) and the Schwarzschild black hole is the special case where the charges $Q_i$ are all zero. The black hole solutions, the associated thermodynamic variables, and the resulting equations of state, are all reviewed in the next two sections. 

The key point in both of these classes of solution, observed in ref.~\cite{Johnson:2019vqf}, is as follows. An examination of the specific heat as temperature varies, $C_V(T)$, shows that it has (for a particular choice of volume) a peak  at some finite $T$, and then a decrease to zero as $T$ grows. {\it This is characteristic of a finite window of available energy states in the system.} For Kerr--AdS (or the three--equal--charge STU--AdS examples, which we will focus on here) the peak's position and height (a measure of the size of our subsystem) is controlled by $J$ (or $Q$). 

Before proceeding, we note that the previous section has also provided a lesson/suggestion about the meaning of the extended gravitational thermodynamic volume, $V$. It has been a puzzle as to what it means in the context of holographically dual field theory, with some suggestions made in the literature\cite{Kastor:2009wy,Johnson:2014yja,Dolan:2014cja,Kastor:2014dra,Caceres:2015vsa,Couch:2016exn}. Here, we see that untangling it from the entropy, combined with trying to make contact with quantum heat engines has opened up a new possibility: It is an effective coupling, somewhat analogous to how the 1D harmonic oscillator's frequency $\omega$ acts as an inverse volume parameter for the thermodynamics, as we reviewed in the previous section. Much of what we do in the rest of this paper will be consistent with this as a natural interpretation, although it would be fruitful to try to prove the correspondence directly in the dual field theory.

\subsection{Kerr--AdS Black Holes}

\label{sec:kerr-black-holes}

For our first exhibit with $C_V\neq0$, let us turn to the Kerr--AdS spacetime, which has metric\cite{Carter:1968ks,Plebanski:1976gy}:
\begin{eqnarray}
ds^2&=&-\frac{\Delta}{\rho^2}\left(dt-\frac{a\sin^2\theta}{\Xi}d\phi\right)^2 +\frac{\rho^2}{\Delta}dR^2+\frac{\rho^2}{\Delta_\theta}d\theta^2+\frac{\Delta_\theta\sin^2\theta}{\rho^2}\left(adt-\frac{R^2+a^2}{\Xi}\right)^2\ ,\nonumber\\
&&{\rm with}\quad\Delta\equiv \frac{(R^2+a^2)(\ell^2+R^2)}{\ell^2}-2mR\ ,\quad \Delta_\theta\equiv1-\frac{a^2}{\ell^2}\cos^2\theta\ , \nonumber\\
&& {\rm and}\quad \rho^2\equiv R^2+a^2\cos^2\theta\ , \quad \Xi\equiv1-\frac{a^2}{\ell^2}\ .
\end{eqnarray}
(Here we are working in four dimensions, for clarity.)
The horizon at $R_+$ is  the largest solution of $\Delta(R_+)=0$. The quantities:
\begin{equation}
\label{eq:physical-kerr}
M=m/\Xi^2\ , \quad  {\rm and}\quad  J=aM
\end{equation}
 are the physical mass and angular momentum, respectively. The entropy is  a quarter of the area of the horizon, $S=\pi(R_+^2+a^2)/\Xi$, but the thermodynamic volume turns out to be independent of~$S$\cite{Cvetic:2010jb}. It is hard to write the equation of state $T(p,V)$ in closed form, but the mass, and a bit of algebra, yields the enthalpy  in the form $H(S,p,J){=}M$, from which $T(S,p,J)$, and $V(S,p,J)$,  are readily derived\cite{Caldarelli:1999xj,Dolan:2011xt}: 
\begin{eqnarray}
H(S,p,J)&=& \frac{1}{2}\sqrt{\frac{\left(S+\frac{8pS^2}{3}\right)^2+4\pi^2\left(1+\frac{8pS}{3}\right)J^2}{\pi S}} \ ,\label{eq:kerr-enthalpy}\\
T(S,p,J)&=&\frac{1}{8\pi H}\left[\left(1+\frac{8pS}{3}\right)\left(1+8pS\right)-4\pi^2\left(\frac{J}{S}\right)^2\right] \ ,\label{eq:kerr-temp} \\
V(S,p,J)&=&\frac{2}{3\pi H}\left[S\left(S+\frac{8pS^2}{3}\right)+2\pi^2 J^2\right]\ .\label{eq:kerr-volume}
\end{eqnarray}

\subsection{STU Black Holes}
\label{sec:stu-black-holes}

The STU--AdS metric is (again, working in four dimensions for presentational clarity)\cite{Sabra:1999ux,Duff:1999gh,Cvetic:1999xp}:
\begin{eqnarray}
	ds^2&=& -(H_1H_2H_3H_4)^{-1/2}f(r)dt^2+(H_1H_2H_3H_4)^{1/2}\left(f(r)^{-1}dr^2+r^2(d\theta^2+\sin^2\theta d\phi^2)\right)\ ,\nonumber\\
	 &&{\rm with}\quad f(r)\equiv1-\frac{2m}{r}+\frac{r^2}{\ell^2} H_1H_2H_3H_4\ ,
\end{eqnarray}
and  the functions $H_i=(1+q_i/r)$ are given in terms of four parameters $q_i$ ($i=1,\cdots,4$). The horizon is at $r=r_+$, where $r_+$ is given by $f(r_+)=0$. This equation determines the parameter $m$ in terms of $q_i$ and $r_+$:
\begin{equation}
\label{eq:m-equation}
m=\frac{r_+}{2}\left(1+\frac{r_+^2}{\ell^2}\prod_iH_i(r_+)\right)\ .
\end{equation}
Then the extended thermodynamics yields\cite{Caceres:2015vsa} the following expressions for the enthalpy $H$, as well as $T, S,V$, and $p$:
\begin{eqnarray}
        H &=& M=m+\frac14\sum_i q_i\ , \label{eq:enthalpy}\\
	T &=& \frac{f^\prime(r_+)}{4\pi}\prod_i H_i^{-1/2}(r_+)\ , \label{eq:temperature}\\
	S &=& \pi \prod_i \sqrt{r_++q_i}\ , \label{eq:entropy}\\
	V &=& \frac{\pi}{3}r_+^3\prod_i H_i(r_+)\sum_j H_j^{-1}(r_+)\ , \label{eq:volume}\\
	p &=& \frac{3}{8\pi G\ell^2}\ . \label{eq:pressure}
\end{eqnarray}
The parameters $q_i$ are related to a family of four physical charges that are given by
   \begin{equation}
   Q_i=\frac{1}{2}\sqrt{q_i(q_i+2m)}\ .
   \label{eq:fixedcharge}
   \end{equation}
In the special case of all four charges being equal, the scalars decouple, and the solution becomes  Reissner--Nordstr\"{o}m--AdS, with $Q=\sum_i Q_i/4$.

 This leads to a nice picture: choices that break the symmetry to give the more general STU holes  are  akin to a sort of ``doping'' process. Degrees of freedom (resulting in a finite energy window) are  added to the system resulting in $C_V\neq0$.  This is somewhat analogous to the sort of doping one might do in a material, resulting in Schottky peaks in the experimental data.

 For simplicity, we will restrict ourselves in this paper to the case of three equal charges, which we will sometimes denote as 3-$Q$ (Reissner--Nordstr\"om is 4-$Q$ in this notation).

 \subsection{Schottky--like peaks, and a Critical point}
 \label{sec:schottky-peaks}
\begin{figure}[h]
\centering
\subfigure[]{\centering\includegraphics[width=0.42\textwidth]{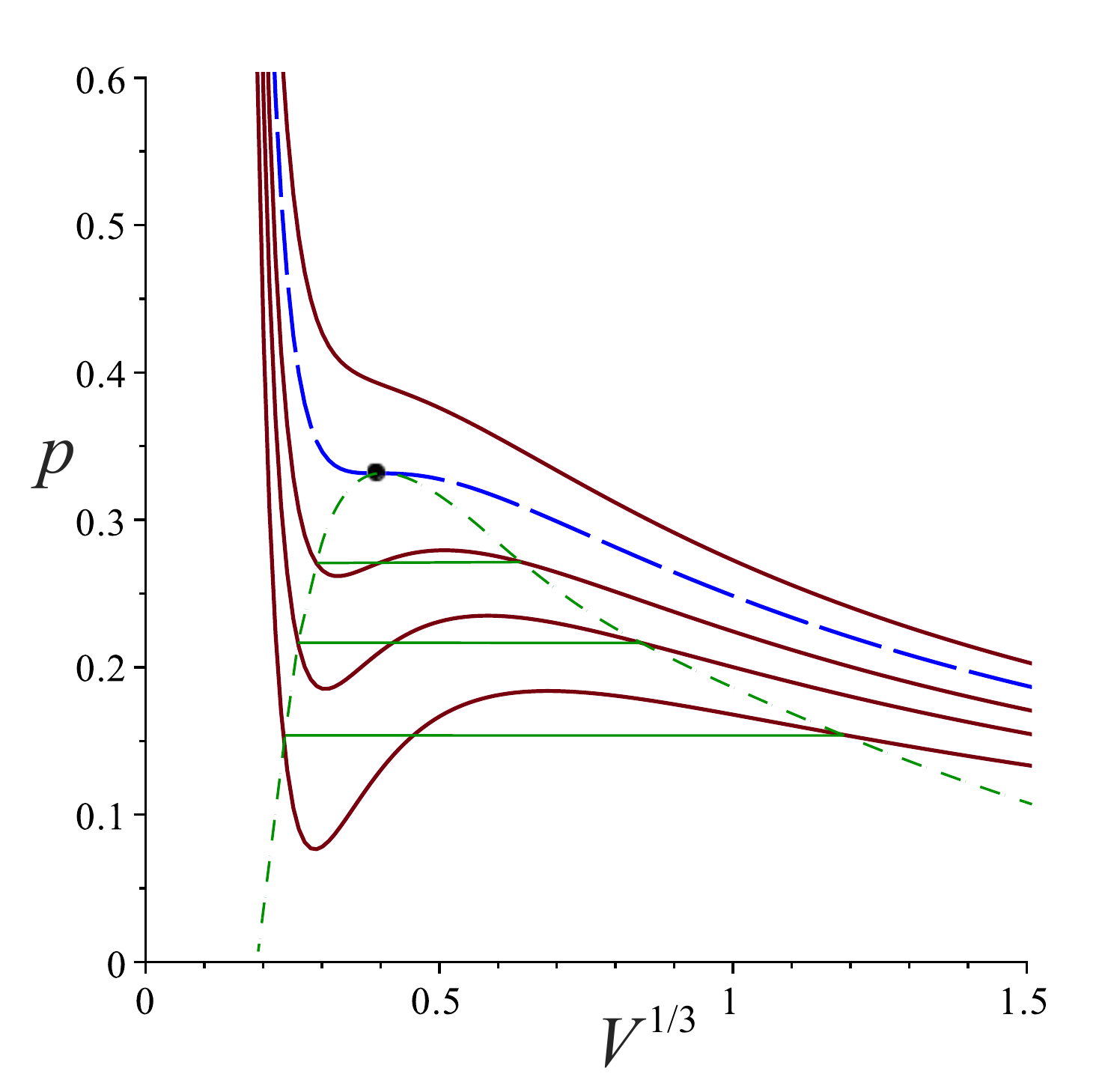}
\label{fig:pv-diagram}}
\subfigure[]{\centering\includegraphics[width=0.50\textwidth]{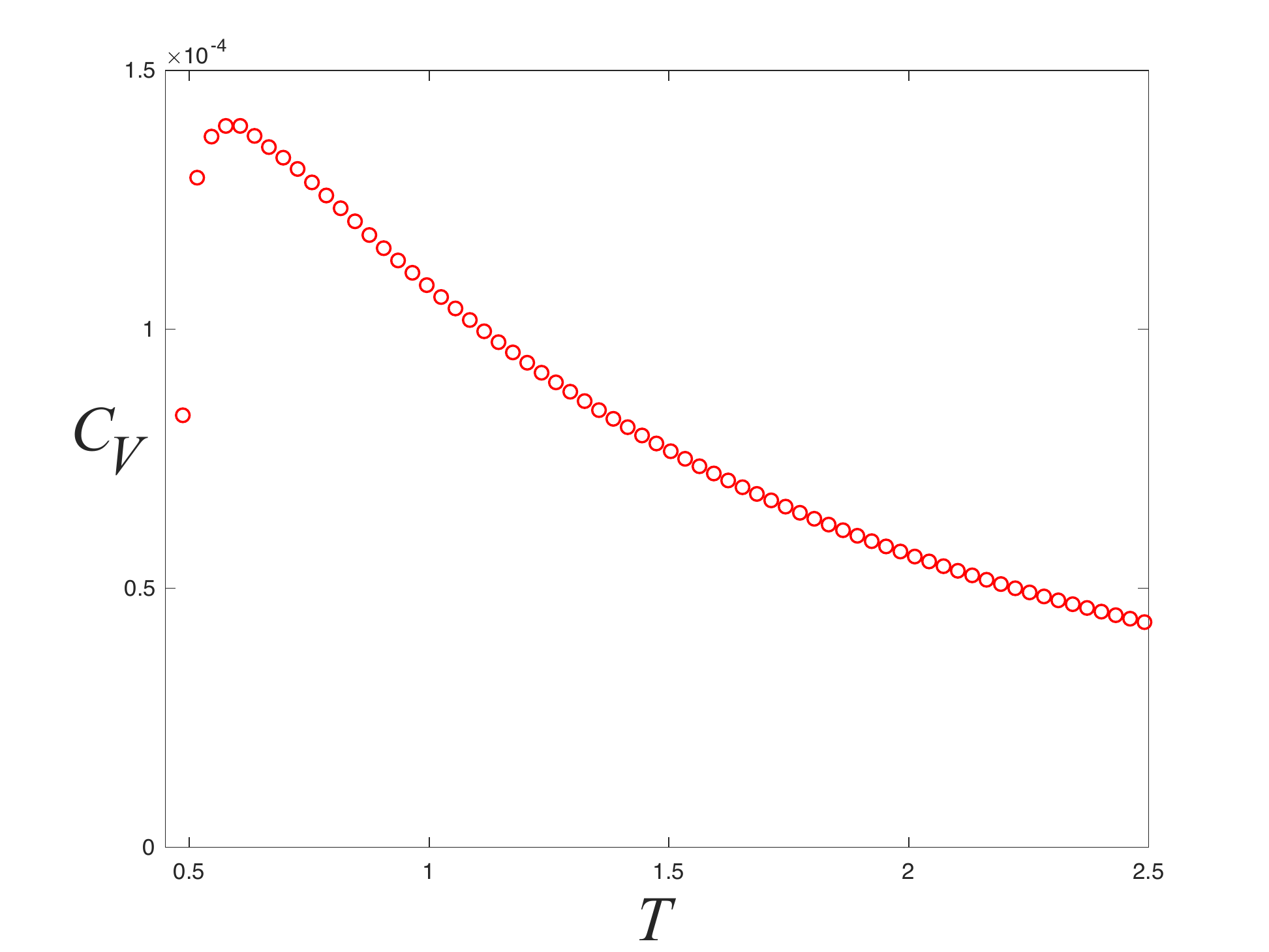}
\label{fig:crit-volume-Cv}}
\caption{(a) A family of isotherms showing the appearance of the critical region as temperature is decreased toward $T_c$ (indicated by the (blue) long--dashed isotherm. There are first order transitions below that temperature. See the text.  (b) The specific heat $C_V(T)$ at   $V\simeq4.166V_c$, showing a Schottky--like peak somewhat above $T_c\simeq0.539$ (this is the 3-$Q$ STU case, with $Q=0.05$).}
\end{figure}

 For both types of black hole, Kerr--AdS and the 3-$Q$ STU black hole, the overall phase structure is  qualitatively similar to the van--der Waals structure discovered for 4-$Q$ in ref.\cite{Chamblin:1999tk}\footnote{See ref.~\cite{Caldarelli:1999xj} for the Kerr--AdS case. These were then all cast into extended thermodynamics language in refs.\cite{Kubiznak:2012wp,Gunasekaran:2012dq,Caceres:2015vsa}.}. Figure~\ref{fig:pv-diagram} displays a family of isotherms showing that there is a critical temperature $T_c$ at which  an unstable (positive slope)  region develops which is in fact excluded by a family of first order transitions (indicated by the green horizontal lines). We will avoid the first order region in all that we do in this paper, for simplicity, but the neighbourhood of the critical point will be of interest to us for two reasons. The first reason is that, as observed in ref.~\cite{Johnson:2019vqf}, the peak of the Schottky--like behaviour is also in the neighbourhood of the critical point (note that the round top of the peak is accessible above $T_c$). See figure~\ref{fig:crit-volume-Cv} for an example of $C_V(T)$. The second reason will emerge in section~\ref{sec:critical-engines}.

\begin{figure}[h]
\centering
\subfigure[]{\centering\includegraphics[width=0.32\textwidth]{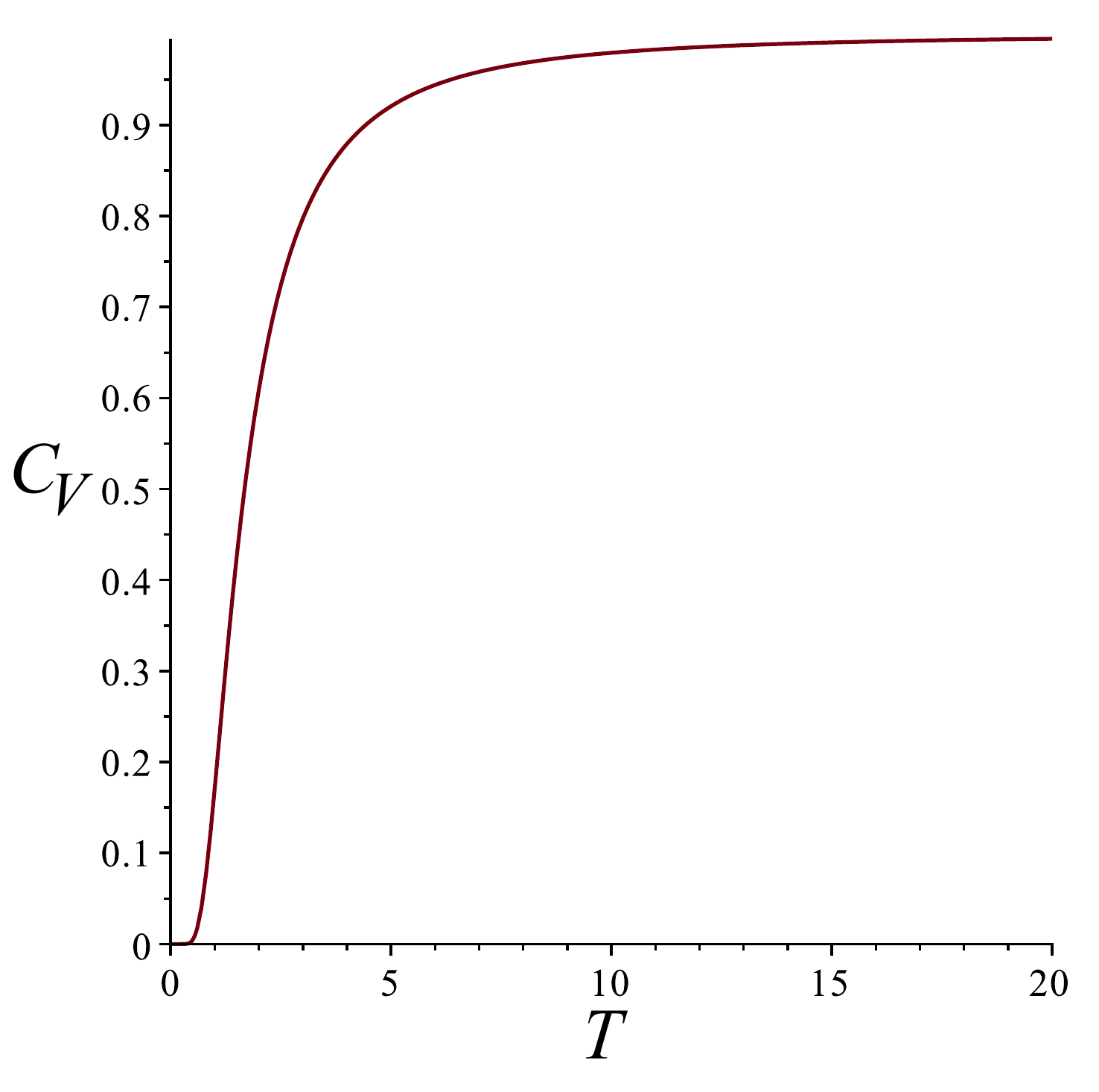}
\label{fig:QHO-Cv}}
\subfigure[]{\centering\includegraphics[width=0.32\textwidth]{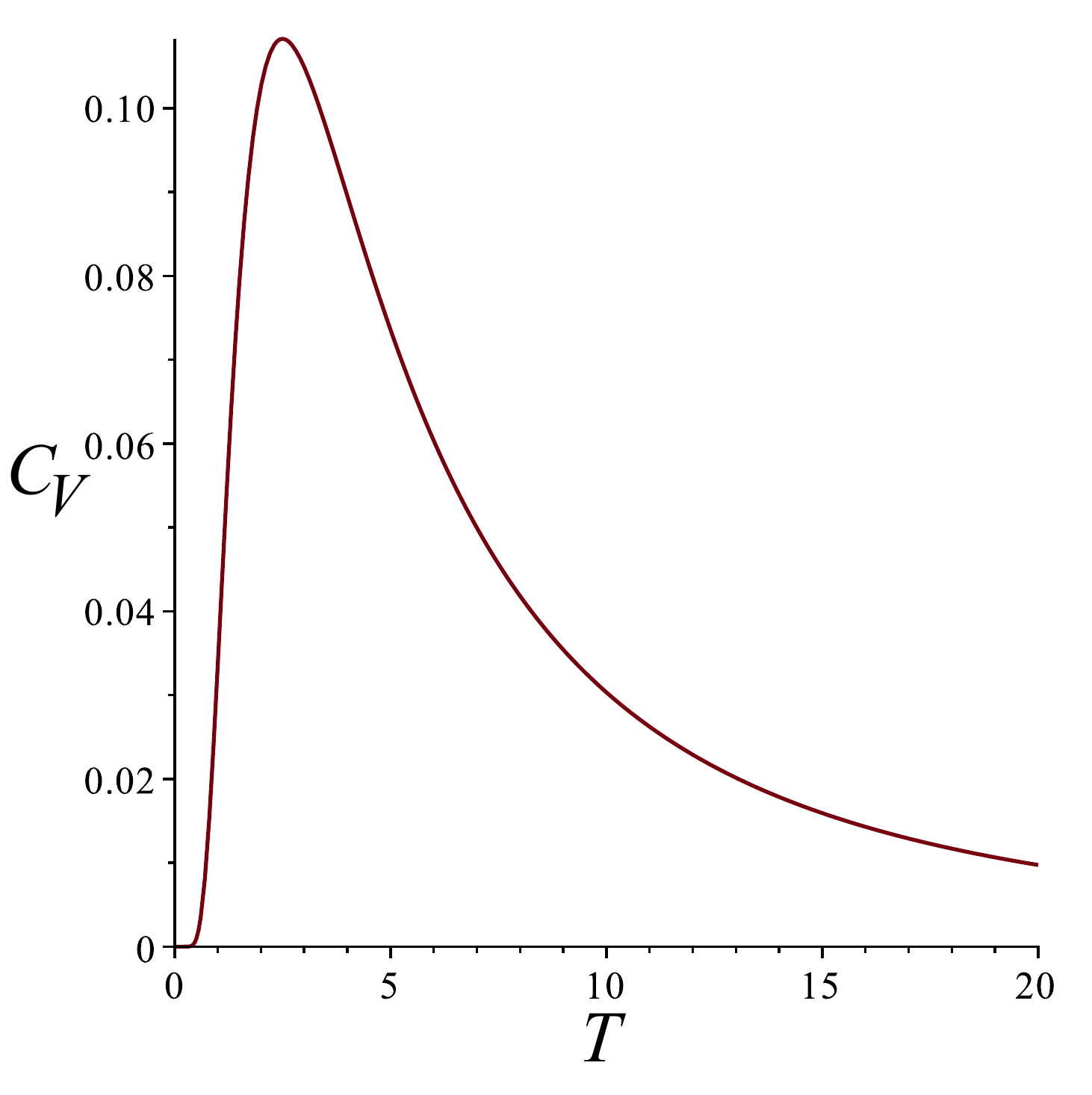}
\label{fig:QHO-truncated-Cv}}
\subfigure[]{\centering\includegraphics[width=0.32\textwidth]{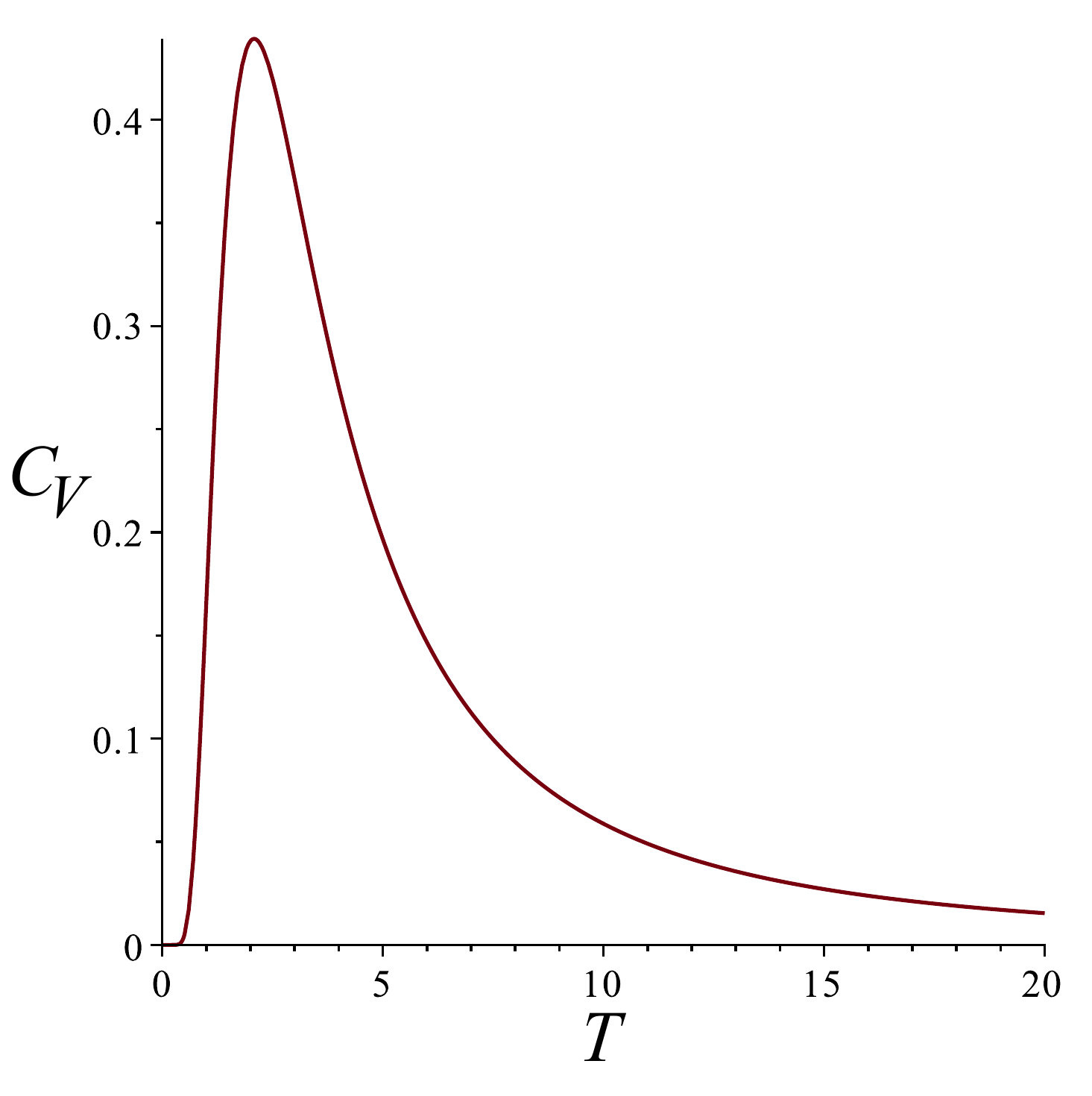}
\label{fig:Schottky-Cv}}
\caption{Three examples of the temperature dependence of the specific heat $C_V(T)$ (a) The  quantum harmonic oscillator, with $\Delta=5$. (b) The quantum harmonic oscillator truncated to $n=5$. Here, $\Delta=1$. (c) A two--level Schottky model, with $\Delta=5$.}
\end{figure}

It is worth pausing here to understand the nature and implications of the peak. It is central to our overall proposal. We do not have a direct path by which we can start with the dual gauge theory, find  an interpretation there of the black hole's thermodynamic volume, and then solve for the reduced model that results when working in the fixed volume ensemble. Whether that is possible or not is left for future research. However $C_V(T)$ is a very specific clue to the results of such a process. The peak is a clear earmark of a finite window of available energies for the effective system. The high temperature regime, which we have a clear sight of well away from the critical region, contains an exponential fall off of the form $\exp(-\Delta/k_BT)$ (where $\Delta$ is a characteristic energy scale of the problem), possibly multiplied by powers of $T$. The low temperature limit, generically obscured by the critical regime here, would have the same form, sending $C_V(T)\to0$ as $T\to0$ as is consistent with Nernst's theorem. Two extremely simple (and highly relevant) models that have specific heats of this form are the Schottky specific heat for (say) a two--level system and a truncated quantum harmonic oscillator. The specific heat of the quantum harmonic oscillator, with energies $E_n=(n+\frac12)\hbar\omega$ is 
\begin{equation}
C_V^{\rm QHO}=k_B\left(\frac{\Delta}{k_BT}\right)^2\frac{{\rm e}^{\beta\Delta}}{(1-{\rm e}^{\beta\Delta})^2}\ , \qquad {\rm where}\quad \beta=\frac{1}{k_B T}\ ,\quad \Delta=\hbar\omega\ .
\end{equation}
This is the classic 1907 model that Einstein used for modelling experimental features of specific heats\cite{Einstein1907}, and is displayed in figure~\ref{fig:QHO-Cv}. 
Notice that  this specific heat has a representation as an infinite series:
\begin{equation}
C_V^{\rm QHO}=k_B\left(\frac{\Delta}{k_BT}\right)^2\sum_{n=1}^\infty n {\rm e}^{-n\beta\Delta}\ ,
\end{equation}
and a truncation to any finite $n$ (effectively turning off the contributions above a certain excitation level) will result in a peak of the form under discussion. See figure~\ref{fig:QHO-truncated-Cv}. The classic Schottky specific heat for a two level system is closely related in form to the above, with (interestingly) a simple sign flip\footnote{The minus signs are suggestive of fermions. This is possibly merely an amusing coincidence.}:
\begin{equation}
C_V^{\rm Schottky}= k_B\left(\frac{\Delta}{k_BT}\right)^2\frac{{\rm e}^{\beta\Delta}}{(1+{\rm e}^{\beta\Delta})^2} =  k_B\left(\frac{\Delta}{k_BT}\right)^2\sum_{n=1}^\infty  (-1)^{n-1} n {\rm e}^{-n\beta\Delta}\ ,
\end{equation}
where here $\Delta$ is the energy gap between the levels. This is depicted in figure~\ref{fig:Schottky-Cv}.

The point is that these analytic forms of the peaks match well to the ``phenomenologically" discovered~\cite{Johnson:2019vqf} peaks (see {\it e.g.} figure~\ref{fig:crit-volume-Cv}). While we do not have a direct derivation of it from first principles, we see that our physics has a spectrum with a  finite energy window, which corresponds to some truncated spectrum (either continuous or discrete), with an associated scale $\Delta$. This is our working substance, and it is fully modelled by the equations of state supplied by the black holes.
 It compares well to the types of simple quantum systems (spins, harmonic oscillators, {\it etc.,}) used in theoretical and experimental studies of quantum heat engines.

\section{Otto Engines from Black Holes}
\label{sec:black-hole-otto}

Now we are ready to construct an Otto cycle. The goal is to locate the $(p,V,T)$ coordinates of the four points of the cycle (1,2,3,4, as shown in figure~\ref{fig:Otto-cycle}), and then compute the efficiency of the engine.  
\begin{wrapfigure}{r}{0.45\textwidth}
\centering
\includegraphics[width=0.45\textwidth]{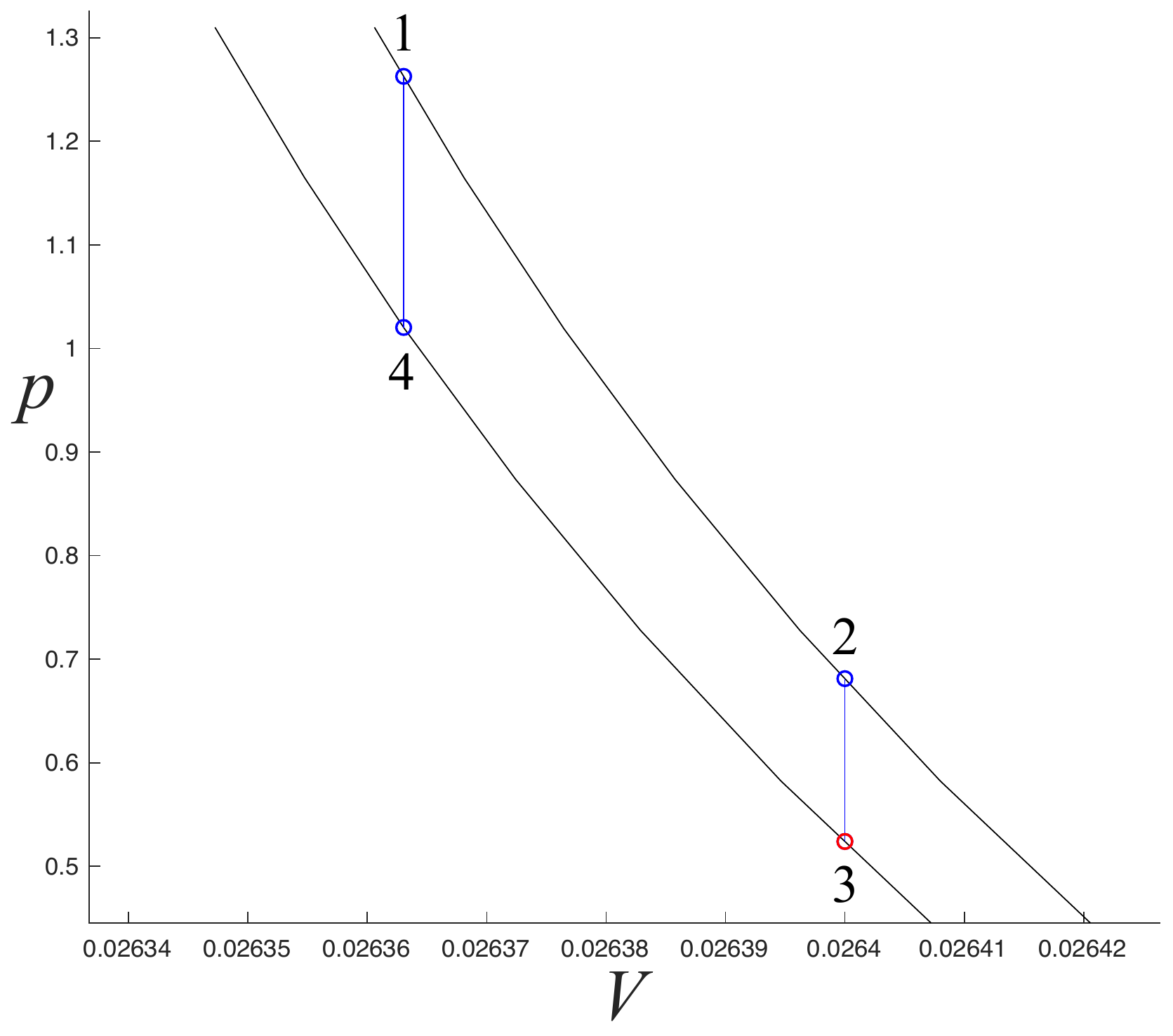}
\caption{\label{fig:Otto-cycle-real} An Otto engine cycle for the 3-$Q$ STU-AdS black hole, where  corner 3 is at $(p_c,V_c)$. Labelling matches that in figure~\ref{fig:Otto-cycle}.}
\end{wrapfigure}
This requires the determination of the heat flows $Q_H$ ($4\to1$) and $Q_C$ ($2\to3$). This can be done by explicitly integrating $C_V(T)dT$ along the isochores. Alternatively, there is a more direct result that can be used because the black hole organizes the thermodynamic variables quite nicely. Since the First Law is $dU=TdS-pdV$, movement along an isochore means that the heat can be simply written as the change in internal energy, and hence:
\begin{equation}
\label{eq:efficiency_otto}
\eta = 1-\frac{U_2-U_3}{U_1-U_4}\ , \quad{\rm where}\quad U= H-pV\ , 
\end{equation}
and $H$ is simply the black hole mass (see eqns.~(\ref{eq:physical-kerr}) and~(\ref{eq:enthalpy})). This equation is the natural analogue of the simple formula written in ref.\cite{Johnson:2016pfa} using masses for cycles where all heat exchanges are on isobars. The second law of thermodynamics ensures that~$\eta\leq\eta_{\rm C}$, the Carnot efficiency (made by constructing a Carnot engine operating between heat baths at the highest ($T_1$) and lowest ($T_3$) temperatures) and is given by:
\begin{equation}
\label{eq:carnot}
\eta_{\rm C}^{\phantom{C}}=1- \frac{T_3}{T_1}\ .
\end{equation}

\subsection{STU Black Hole Otto}
 In practice, since we do not have closed forms for the thermodynamic quantities, our exact formula cannot be used to avoid numerical methods. So, as described in ref.~\cite{Johnson:2019vqf}, once a region of interest in state space has been identified, an $n\times n$ grid of points (typically for $n=100$) was chosen (in the $(p,V)$ plane), over which all the state functions of interest ($T$, $S$, $U$, {\it etc.,}) were computed numerically. The results were then stored for later data mining when analyzing the heat engines. (We used both Maple and MatLab to do this.) 

The strategy for proceeding was to pick the coordinates of a starting corner ($(p_3,V_3)$ say, see figure~\ref{fig:Otto-cycle}).  The value of the state functions such as $S,T, H$, or $U$ can then be mined from the numerical data.  Of course $V_2=V_3$ since corner 2 is on an isochore. The other two choices to make  in order to define the cycle are the value of $p_2$, and the value of $V_4=V_1$. The pressures $p_1$ and $p_4$ follow from the fact that they lie on adiabats connecting to the corners 3 and 2, and so they must be determined, again using the numerical data since  adiabats are not known in closed form. Now  all corners have their coordinates determined, and so  state variables (such as $T,H$ or $U$) can then be mined from the data.

Figure~\ref{fig:Otto-cycle-real} is an example of an Otto cycle constructed in this way using the 3-$Q$ black hole (with $Q=0.05$). We chose to place the corner $(p_3,V_3)$  at the critical point, $(p_c\simeq0.5240,V_c\simeq0.026400)$. We chose $p_2\simeq0.6812$ and $V_1\simeq0.026363$. (The extra digits on volumes are a reflection of the fact that the adiabats are just slightly off being the vertical lines of the $C_V=0$ case for this small choice of $Q$.)  The efficiency of this prototype Otto cycle turns out to be $\eta=0.2625$, while the Carnot efficiency  is $\eta_{\rm C}^{\phantom{C}}\simeq0.3387$, setting the upper bound, by the second law of thermodynamics.

\subsection{Kerr Black Hole Otto}
\label{sec:kerr-otto}
\begin{wrapfigure}{r}{0.45\textwidth}
\centering
\includegraphics[width=0.45\textwidth]{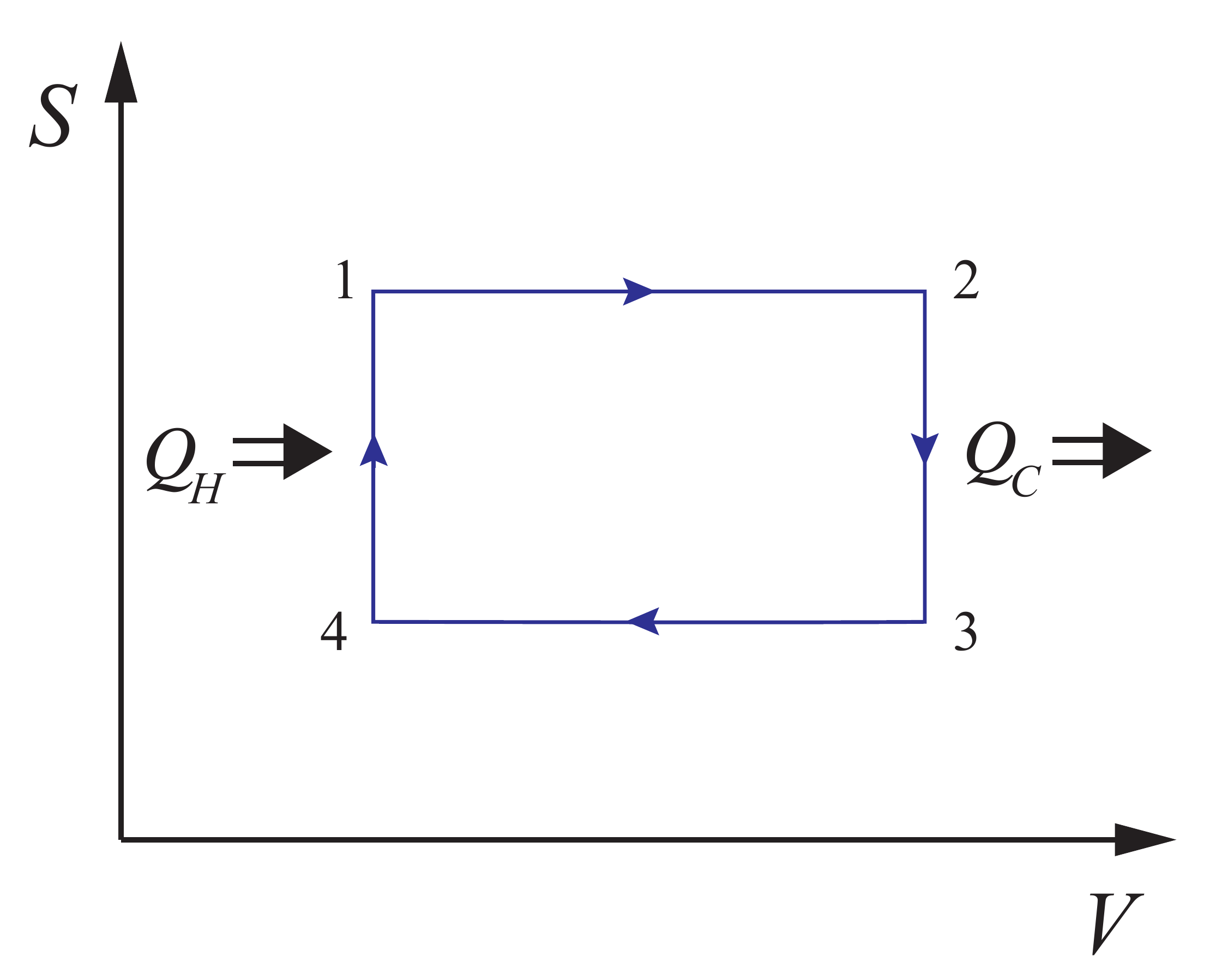}
\caption{\label{fig:otto-rectangle} The Otto engine cycle in the $(S,V)$ plane. Labelling matches that in figure~\ref{fig:Otto-cycle}.}
\end{wrapfigure}

Fortunately, some analytic progress on our Otto engine can be made for the Kerr black hole.  The expression for the volume $V$ in equation~(\ref{eq:kerr-volume}) can be used to solve for  $p$ in terms of $(V,S,J)$ and so~$p$ can be eliminated from $H$ to give\cite{Dolan:2011xt}:
\begin{equation}
\label{eq:kerr-enthalpy-sv}
H(S,V,J)=\frac{1}{2}\sqrt{
\left\{\left(\frac{3V}{4\pi}\right)^2 -\left(\frac{S}{\pi}\right)^3\right\}^{-1}
\left[ \left\{\left(\frac{S}{\pi}\right)^2+2J^2 \right\}^2-  \left(\frac{S}{\pi}\right)^2\left\{\left(\frac{S}{\pi}\right)^2+4J^2 \right\} \right]
}
\end{equation}
Then, $U=H-pV$ can be constructed and written in terms of its natural variables as\cite{Dolan:2011xt}:
\begin{equation}
\label{eq:kerr-energy}
U(S,V, J) = \left(\frac{\pi}{S}\right)^3\left[\left(\frac{3V}{4\pi}\right)\left\{ \frac12\left(\frac{S}{\pi}\right)^2+J^2\right\}-J^2\left\{\left(\frac{3V}{4\pi}\right)^2-
\left(\frac{S}{\pi}\right)^3\right\}^\frac12\right]\ .
\end{equation}
From this $T(S,V,J)=(\partial U/\partial S)_V$ can be written directly, and it is:
\begin{equation}
\label{eq:kerr-temperature}
T
= \frac{3J^2}{\pi}\left(\frac{\pi}{S}\right)^4
\left[ 
\left\{\left(\frac{3V}{4\pi}\right)^2 -\frac12\left(\frac{S}{\pi}\right)^3\right\}
\left\{\left(\frac{3V}{4\pi}\right)^2 -\left(\frac{S}{\pi}\right)^3\right\}^{-\frac12} 
-\left(\frac{3V}{4\pi}\right) 
\left\{1+\frac{3}{18J^2}\left(\frac{S}{\pi}\right)^2\right\} 
\right]    \ .
\end{equation}
This is an extremely convenient pair of thermodynamic functions to write, and in terms of $S$ and~$V$ especially, since the Otto cycle is indeed built from adiabats and isochores and so is represented in an $(S,V)$ diagram as a rectangle (see figure~\ref{fig:otto-rectangle}). We need only choose, say, $(S_3,V_3)$
 and $(S_1,V_1)$ and we can write the efficiency down fully by substituting the values into equation~(\ref{eq:kerr-energy}) to get the~$U_i$, ($i=1,2,3,4$), and then computing equation~(\ref{eq:efficiency_otto}).
 The Carnot efficiency bound~(\ref{eq:carnot}) for that engine is also easily computed  by using the chosen values at corners 1 and 3 to compute $T_1$ and $T_3$. In either case, the expressions for $\eta$ and $\eta_{\rm C}$ are readily written down and are too large to write usefully in single expressions. 
 
 Another natural plane to work in is the $p{-}S$ plane, since expressions~(\ref{eq:kerr-enthalpy}),~(\ref{eq:kerr-volume}),~and~(\ref{eq:kerr-temp}) are written naturally with that dependence. Again, analytic expressions can be written for the efficiencies using these formulae, albeit too long to write usefully here.
 
There is a tradeoff, however. With both choices of independent variables ($(S,V)$ or $(p,S)$), picking coordinates of the corners such that the equations of state are satisfied becomes a delicate matter. This is traceable to our central feature---the fact that $C_V$ is very small. So it requires some care to ensure that the corners can be connected by adiabats, and this again requires the equations of state to be solved numerically.

For stability, it turned out again that it was convenient to do the numerical  work in the  $p{-}V$ plane (by eliminating $S$ from equations~(\ref{eq:kerr-enthalpy}),~(\ref{eq:kerr-volume}),~and~(\ref{eq:kerr-temp}) to obtain $T(p,V)$ and $S(p,V)$) and using the scheme of the previous section. In this way, we  found solutions very similar in structure to that seen in figure~\ref{fig:Otto-cycle-real}, extracting $\eta$ and $\eta_{\rm C}^{\phantom{C}}$. We will discuss more of this in the next section.

Having the form of $U(S,V)$ explicitly~(\ref{eq:kerr-energy}) means that we can use the exact formula~(\ref{eq:efficiency_otto}) to deduce some important features of our engine.  We can write $U= v[1/2s+J^2(1-\alpha)/s^3)]$, where $s=S/\pi$ and $v=3V/4\pi$ and $\alpha=\sqrt{1-s^3/v^2}$, which vanishes when $J=0$. The heat $Q_C=U_2-U_3$ at the  isochore $V_2$ is simply  $Q_C=v_2[1/2s_2-1/2s_3)+J^2K_2]$, where $K_2$ comes from evaluating the difference of the order $J^2$ terms. Similarly $Q_H=v_1[1/2s_1-1/2s_4)+J^2K_1]$, where $K_1>K_2$. Since $s_1=s_2$ and $s_1=s_4$, we can divide by an overall constant $\beta=(s_4-s_1)/2s_1s_4$ to write the ratio of heats as $Q_C/Q_H=V_2/V_1(1-J^2(K_1-K_2)/\beta+\cdots)$, where we treat $J$ as small. At $J=0$, adiabats return to being vertical and $V_1\to V_2$, the engine has zero area and $\eta\to0$.  The difference $K_1{-}K_2$ vanishes at $J=0$ and must increase with the ratio $V_1/V_2$ in such a way that $\eta$ must remain positive (lest we violate the first law), and the rate at which it increases depends upon $J$, since $J$ controls the shape of the adiabats. A reasonable guess as to the dependence  that satisfies these conditions at this order is $K_1{-}K_2=(\beta/J^2)[1-(V_1/V_2)^\gamma]$, where $\gamma>1$. This gives (in the small $J$ regime where we expect to make contact with a quantum heat engine):
\begin{equation}
\label{eq:otto-efficiency}
\eta=1-\left(\frac{V_1}{V_2}\right)^{\gamma-1}\left(1 + O(J^2)\right)\ .
\end{equation}
(We checked explicitly that this power law behaviour for $1{-}\eta$ is indeed borne out in fully (numerically) solved examples for small $J$.)
This  form of the efficiency is  the classic form for the traditional Otto engine, but additionally, it agrees with the form for certain simple quantum Otto engines as well, for a variety of simple working substances\cite{PhysRevE.79.041129}.  (For the 1D harmonic oscillator example reviewed in section~\ref{sec:quantum-heat-engines}, $\gamma=2$.) Here, the constant $\gamma$ is the effective adiabatic exponent $C_p/C_V$. It is an {\it effective} adiabatic exponent  because while we expect a finite $\gamma$ because now $C_V\neq0$,   since both~$C_p$ and $C_V$ vary over the $p{-}V$ plane (as independent functions), and hence~$\gamma$ varies. In our case~$\gamma$ is very large (of order $10^3$ for the types of engines we built in the numerical examples discussed below), reflecting that we have chosen to have a very small number of degrees of freedom. 

While equation~(\ref{eq:otto-efficiency}) is a satisfying form for the outcome, this result also points to a limiting feature of our examples chosen thus far. We have worked with $J$ small, in order to be in a regime where we studying the thermodynamics of  a ``small" subsector of the usual large $N$ dynamics gravity gives access to. On the other hand, $J$ also controls the degree to which we can separate adiabats from isochores, so making it small gives us a large adiabatic exponent, which is not representative of most working substances. (A similar story is likely true for the STU Otto engines, controlled by~$Q$.) Were the goal to simply reproduce efficiency formulae, we would stop at this point. The larger goal, however, as stated in the introduction, is to find a setting where we might be able to usefully interrogate a range of issues concerning quantum heat engines, independently of the size of $\gamma$, and the proposal is that we have.

\section{Running Near Criticality}
\label{sec:critical-engines}
With $J$ (or $Q$) available as a free parameter, it is an interesting question as to how its value affects the Otto engine's performance. Allied to  this question is the issue of whether the neighbourhood of the critical point can have an effect on the efficiency. This matter arose in the  quantum heat engine literature in the context of finding schemes by which the power of a heat engine might be enhanced, while retaining high efficiency. It was suggested\cite{2016NatCo...711895C} that if the working substance had the right kind of critical point, an enhancement could be achieved, principally because the specific heat gets an enhancement  at such a critical point. In fact, this was tested in the holographic heat context in ref.\cite{Johnson:2017hxu} and shown to be qualitatively correct, although the heat engine there was not of the Otto type we are considering here. So it is interesting to revisit that scenario now that we have Otto engines. We do {\it not} expect an enhancement here, however. This is because the critical point for our working substance (modelled by the black hole) gives a divergence in $C_p$, not $C_V$, and it is the latter that is in play for Otto engines. So with no interesting features like a critical point for the constant volume degrees of freedom, we'd expect that there should be no $J$--dependence for the ratio, and we will see that this expectation is borne out. 

Mimicing the setup of ref.\cite{Johnson:2017hxu}, the idea is to keep the engine cycle always adjacent to the critical point, and then to study the dependence of the efficiency $\eta$ on $J$ (or $Q$). It is natural to compare to the Carnot efficiency of that same engine, and so   we study the ratio $\eta/\eta_{\rm C}^{\phantom{C}}$. For simplicity, and because there was much greater numerical control of the equations of state, we will focus on the case of the Kerr Otto engine, but our results for the STU Otto engines were similar.
The position of the critical point is given, for small $J$ by\cite{Altamirano:2014tva}:
 \begin{equation}
 \label{eq:critical-point}
 p_c=\frac{1}{12\pi (90)^{1/2} }\frac{1}{J}\ , \quad S_c=\pi (90)^{1/4}J\ , \quad V_c= \frac{4\pi}{3}(90)^{3/4} J^{3/2}\ , \quad T_c=\frac{(90)^{3/4}}{225\pi} \frac{1}{J^{1/2}}\ .
 \end{equation}
  To make the comparison at different values of the $J$, we chose a region of the plane of a size that scaled with the location with the critical point ({\it e.g.,} the lower bound on $p$ and $S$ was $p_c/4$ and $S_c/4$, respectively, and upper bound was $2p_c$ and $6S_c$). We numerically solved the equations of state there. Picking the corners $(p_3,V_3)=(p_c,V_c)$ and $(p_2,V_2)=(3p_c/2,V_c)$ (see labelling in figure~\ref{fig:Otto-cycle-real} or figure~\ref{fig:Otto-cycle}) we numerically solved the adiabat from there to find the largest $p_1$ that would fit into the chosen region. We then found the corresponding $V_1=V_4$, determining the termination point when solving the adiabat that runs from corner~3. Knowledge of the coordinates of corners~1 and~3 were enough to determine the temperatures there and hence $\eta_{\rm C}^{\phantom{C}}$.  
\begin{figure}[h]
\centering
\subfigure[]{\centering\includegraphics[width=0.46\textwidth]{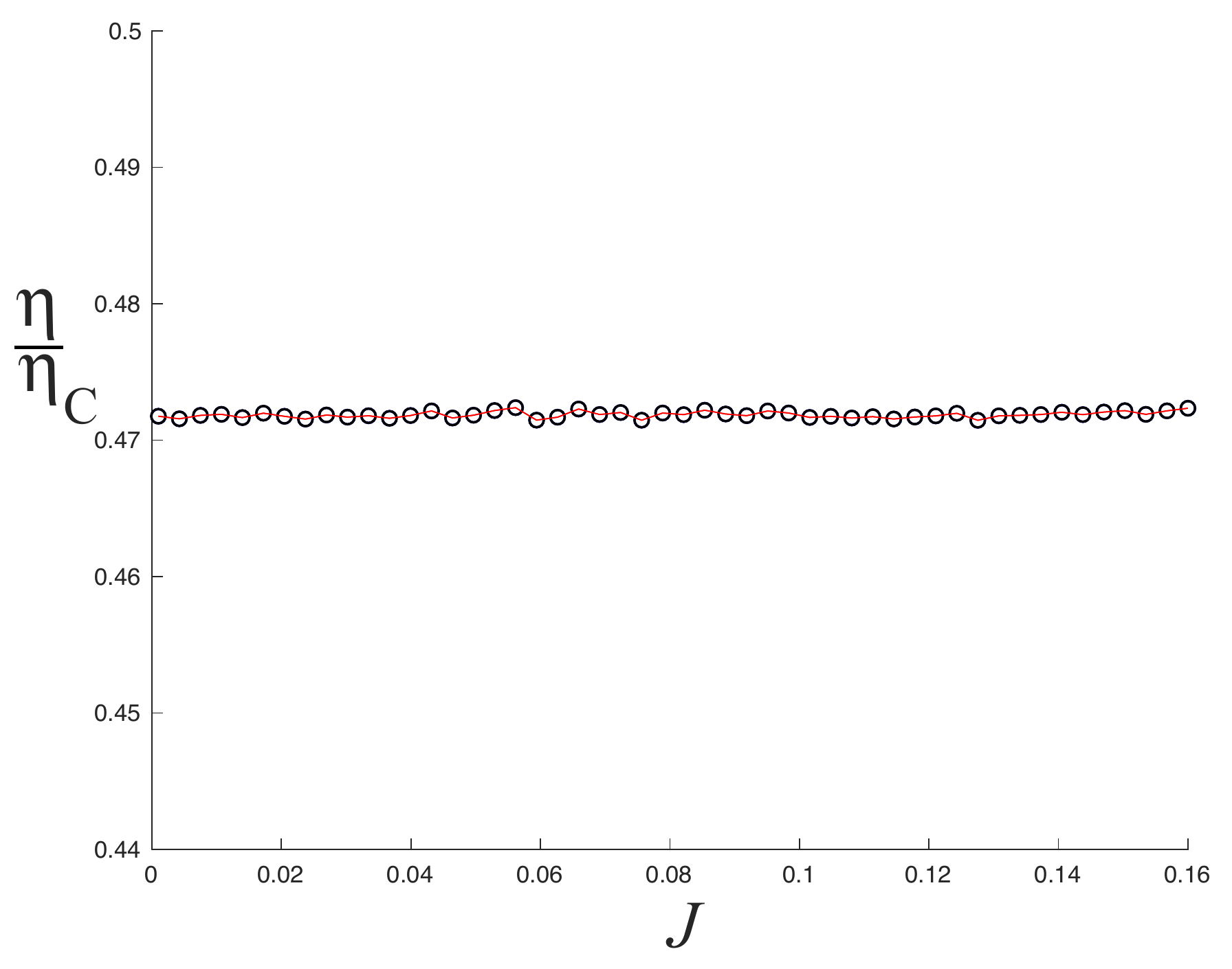}
\label{fig:efficiency-ratio-vs-J}}\hskip1cm
\subfigure[]{\centering\includegraphics[width=0.46\textwidth]{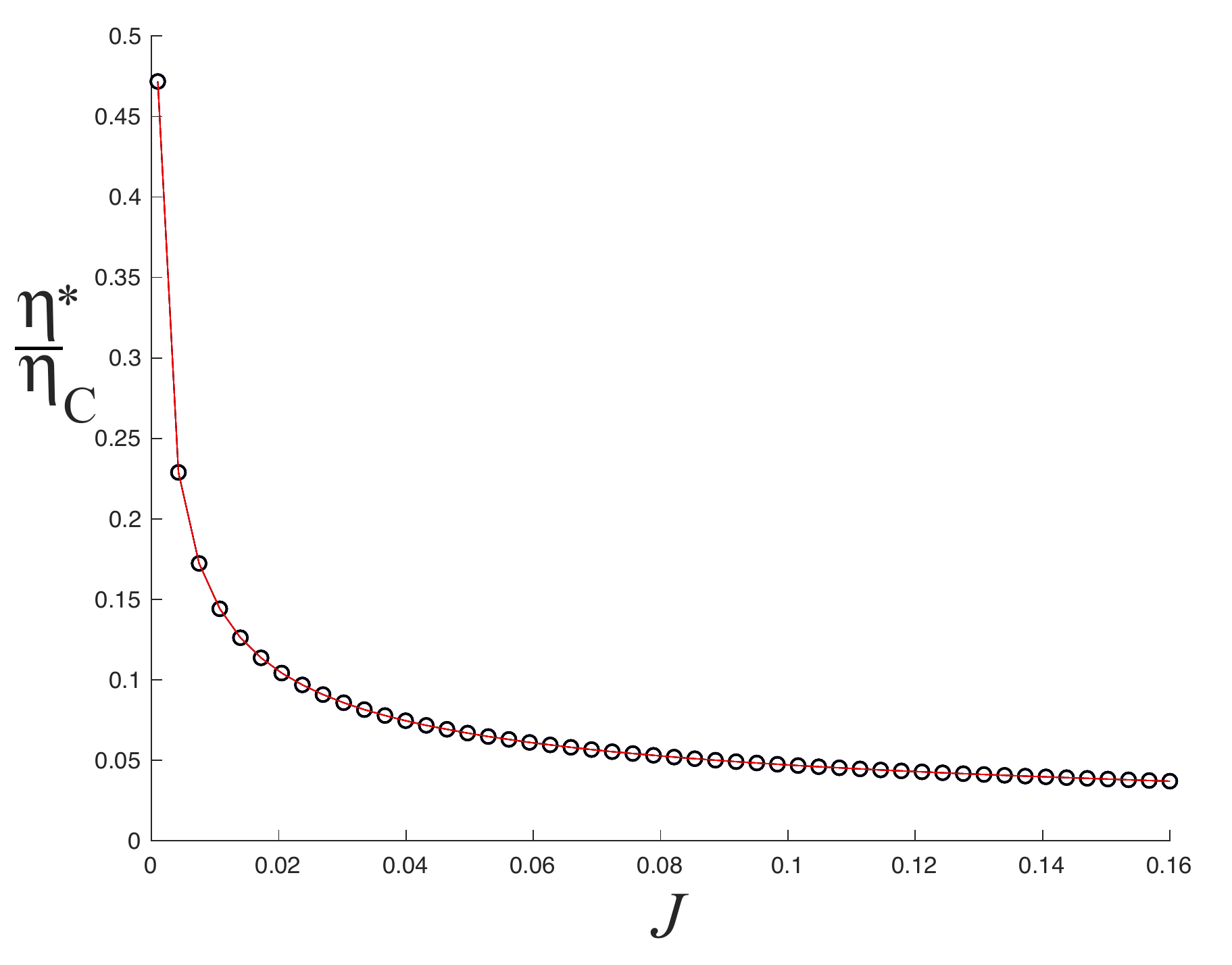}
\label{fig:scaled-efficiency-ratio-vs-J}}
\caption{ (a) The ratio of the efficiency of the (Kerr black hole) Otto engine to its Carnot efficiency. (b) The ratio of the rescaled (see text) efficiency of the (Kerr black hole) Otto engine to its Carnot efficiency.}
\end{figure}
Our results are in figure~\ref{fig:efficiency-ratio-vs-J} and~\ref{fig:scaled-efficiency-ratio-vs-J}, and the heat flows and work performed  were computed by both direct integration along each branch of the cycle, and by computing differences in $U$, as a check of our methods.
 We see in figure~\ref{fig:efficiency-ratio-vs-J} that $\eta/\eta_C^{\phantom C}$ retains a constant (within numerical error) value of $\sim0.472$  as $J$ varies over the range $(0.001,0.16)$.\footnote{Not far beyond this range, the small $J$ validity of  equation~(\ref{eq:critical-point}) for the location of the critical point began to break down, and using it began to insert the heat engine into  the critical region containing first order transitions, with unreliable results.} This confirms the expectation expressed above.

An engine of a given design can be made more efficient by simply making it larger  {\it i.e.,} allowing it to occupy a larger area in the $p{-}V$ plane. Our choice of how we placed the engine near the critical point, described above, amounts to a work that scales as $W\sim J^{1/2}$ (this follows from the scaling of $p_c$ and $V_c$ and we verified this by computing the work at each $J$). We can have a different measure of the relative performance if  instead we rescale the efficiencies such that they are all normalized to the same fixed work $W^*$, giving   $\eta^*=\eta \,\, (W^*/W)$, where $W^*$ was chosen as the work at $J=0.001$. In this way  the scheme becomes closer to that in ref.\cite{Johnson:2017hxu}, where the work was held constant when comparing different  charges. The ratio  $\eta^*/\eta_C^{\phantom C}$  would be expected to fall as $1/J^{1/2}$, and this is confirmed by the result   shown in figure~\ref{fig:scaled-efficiency-ratio-vs-J}.

\section{Brayton and Diesel Cycles}
\label{sec:brayton-diesel}
\begin{figure}[h]
\centering
\subfigure[]{\centering\includegraphics[width=0.46\textwidth]{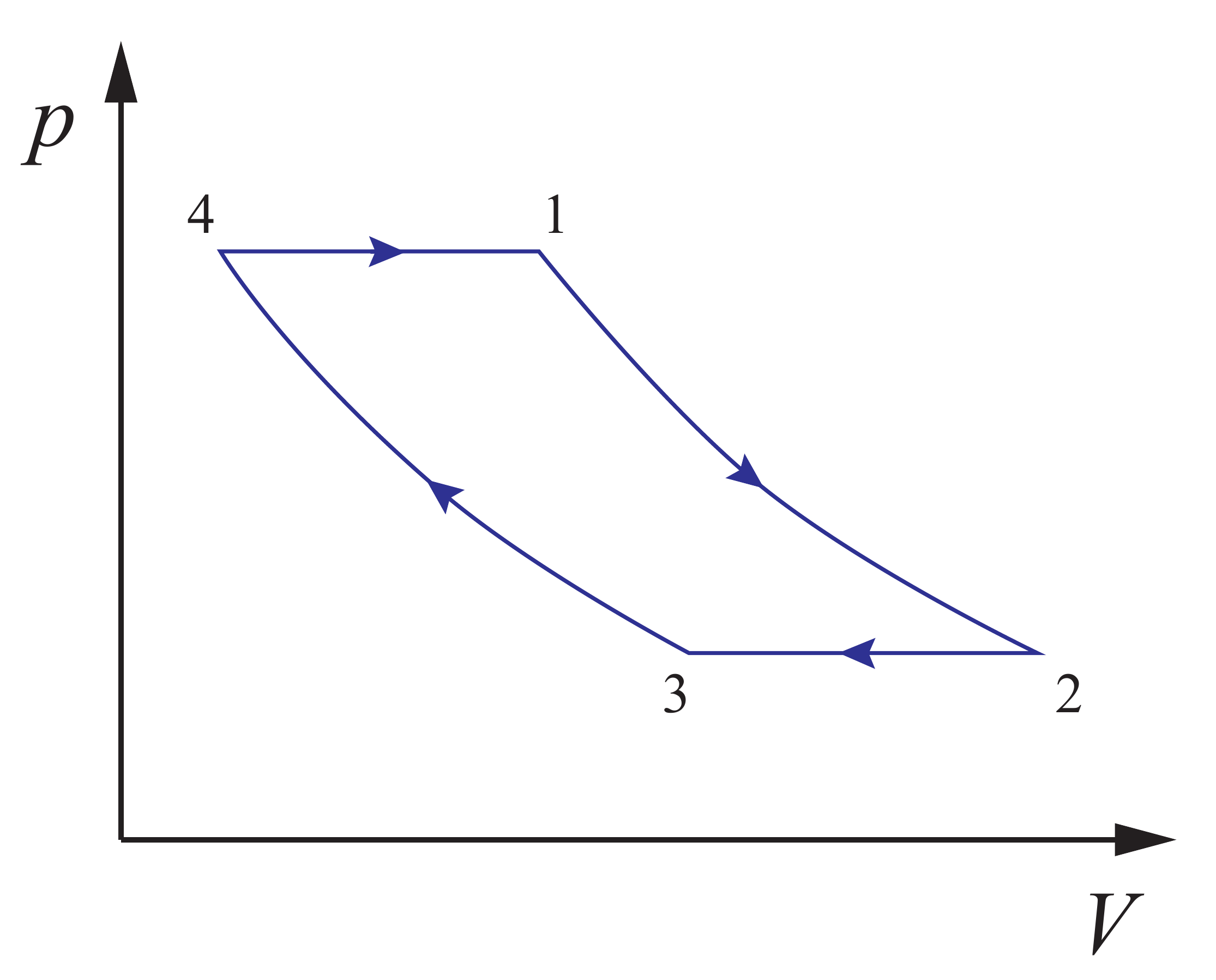}
\label{fig:brayton-cycle}}\hskip1cm
\subfigure[]{\centering\includegraphics[width=0.46\textwidth]{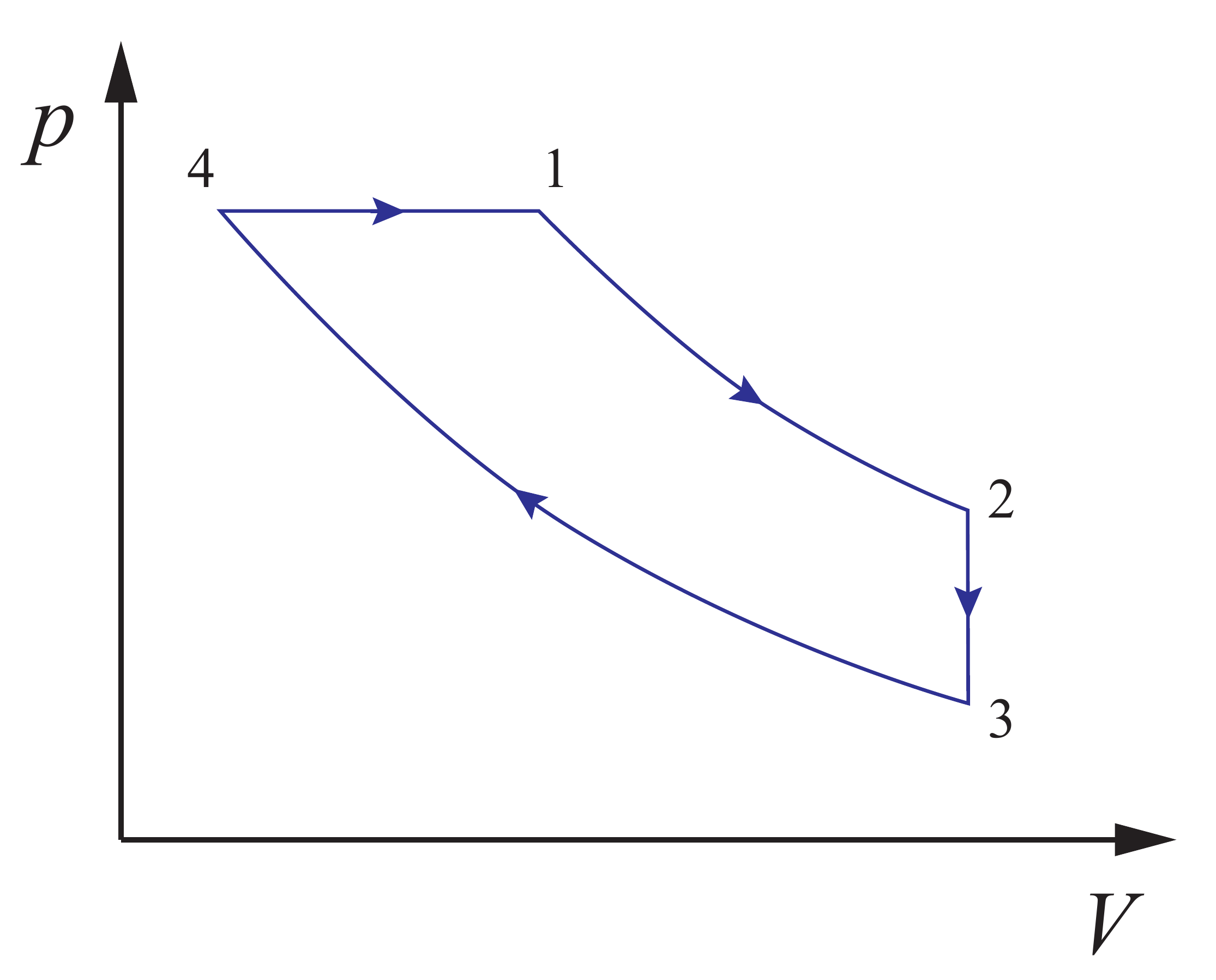}
\label{fig:diesel-cycle}}
\caption{ (a) The Brayton cycle, made from two isobars and two adiabats. (b) The Diesel cycle made from two adiabats, an isobar, and an isochore.}
\end{figure}

It is straightforward  to construct other classic engine cycles. The two most obvious to consider are the Brayton cycle and the Diesel cycle, sketched in figures~\ref{fig:brayton-cycle} and~\ref{fig:diesel-cycle} respectively. 
It is interesting to consider black hole versions of these cycles  in the  context of this paper because there have been quantum heat engine constructions that use these cycles\footnote{We thank Nicole Yunger Halpern for a helpful comment about the literature in this regard.} (see the discussion in refs.~\cite{PhysRevE.76.031105,PhysRevE.79.041129}). An appealing feature of these engines is that we again can use the fact that the gravitational thermodynamics so readily supplies the enthalpy~$H$ as the mass $M$ of the black hole, from which the volume is readily derived and so $U{=}H{-}pV$ is straightforward to compute, as mentioned in earlier sections. Since the first law is $dH{=}TdS{+}Vdp$ or $dU{=}TdS{-}pdV$ and we have heat exchanges on either isobars or isochores, all heat exchanges can be written as either $H$ differences or $U$ differences. So the efficiencies are readily written  as:
\begin{equation}
\label{eq:efficiencies}
\eta_{\rm B}^{\phantom B}= 1-\frac{H_2-H_3}{H_1-H_4}\ , \quad
{\rm and}
\quad
\eta_{\rm D}^{\phantom D}= 1-\frac{U_2-U_3}{H_1-H_4}\ .
\end{equation}
The first is the exact formula of ref.\cite{Johnson:2016pfa} and they both are natural counterparts to equation~(\ref{eq:efficiency_otto}).\footnote{The cycle made by exchanging the position of the isobar and icochore in Diesel is easy to include too, with  the efficiency $\eta=1-(H_2-H_3)/(U_1-U_4)$.} 
\begin{wrapfigure}{r}{0.45\textwidth}
\centering
\includegraphics[width=0.45\textwidth]{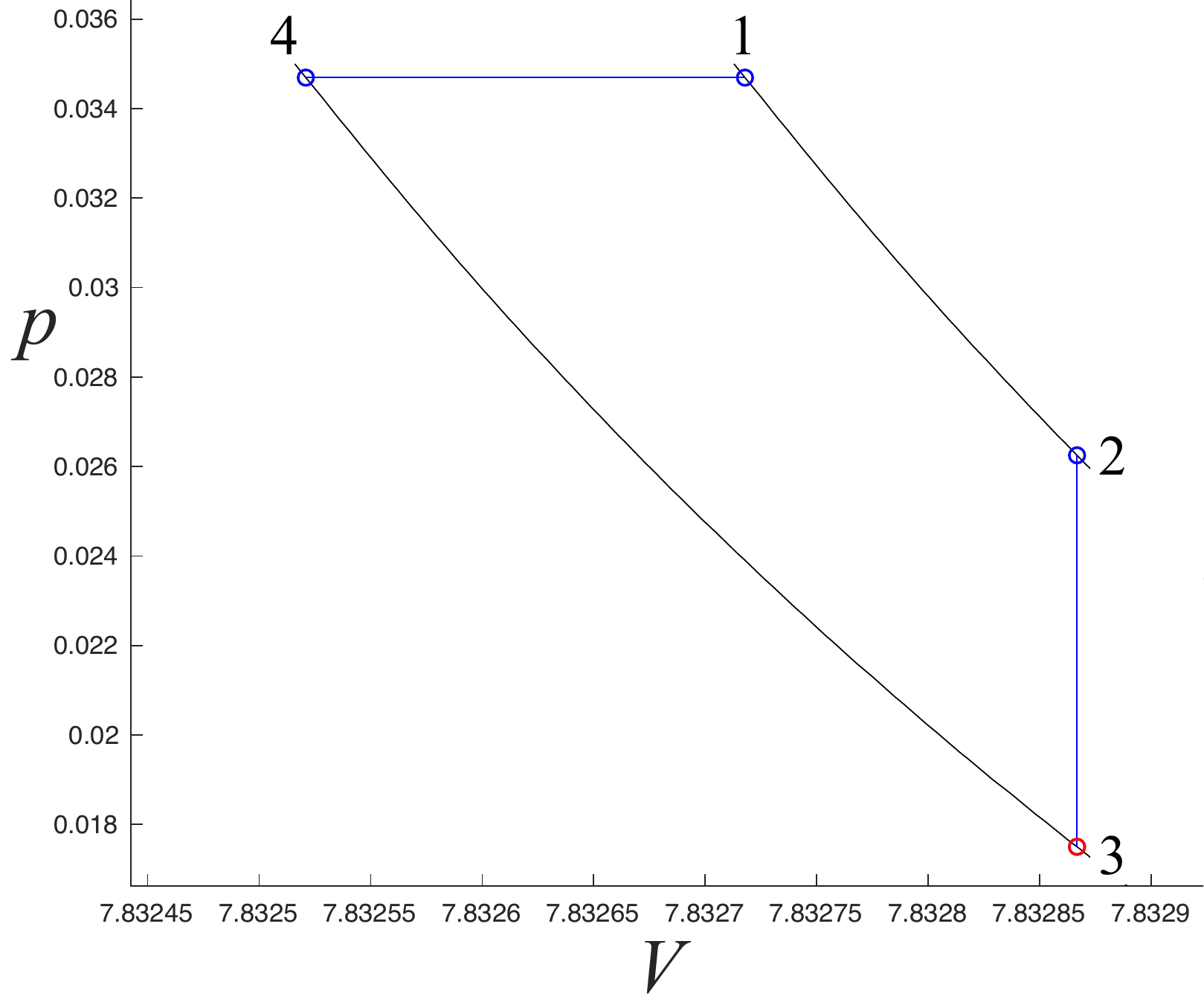}
\caption{\label{fig:Diesel-cycle-real} A Diesel engine cycle for the $J=0.16$ Kerr-AdS black hole, where  corner 3 is at $(p_c,V_c)$. Labelling matches that in figure~\ref{fig:diesel-cycle}.}
\end{wrapfigure}
Using these  it was straightforward to construct some examples using as working substance the Kerr and STU black holes presented here, using techniques similar to those we employed in section~\ref{sec:black-hole-otto}. See figure~\ref{fig:Diesel-cycle-real} for an example of a Diesel engine made from a Kerr black hole, computed at $J=0.16$, with $\eta\simeq0.2216$ and $\eta_{\rm C}^{\phantom{C}}\simeq0.2903$.\footnote{We also studied an analogue of the scheme of section~\ref{sec:critical-engines} where the behaviour of the efficiency of Deisel as a function of $J$ or $Q$ was examined. We found curves similar to those presented for Otto.}

While these were interesting to construct (at least as proof of concept) there is a key reason why they are perhaps less compelling than our Otto black hole engines: For the isobaric components the heat exchanges involve the large number of degrees of freedom (scaling like a positive power of~$N$) that can be excited in fixed pressure processes. In other words, the dual quantum system is the unrestricted large $N$ gauge theory, not the small window afforded by the fixed volume processes (see subsection~\ref{sec:schottky-peaks}). Therefore the thermodynamics involves a rather classical system on those branches. This makes less compelling any argument that these are quantum heat engines (in the sense alluded to in subsection~\ref{sec:synthesis}). At best, perhaps Diesel is  a hybrid sort of system, with a small system on the isochoric part and large on the isobaric part. It is possible that such a large$+$small hybrid  may well be of interest, maybe even  experimentally, in the quantum heat engine context, so it is worth observing that they can arise naturally as black hole models.

\section{Discussion}
\label{sec:discussion}

Two key features formed the foundation of this paper's proposal. The first is that extended gravitational thermodynamics provides solvable models of equations of state for systems that have known (for $\Lambda<0$) underlying microscopic quantum descriptions, because of holographic duality. The second is that constant volume processes have a highly reduced window of energy excitations\cite{Johnson:2019vqf}, suggesting that the dynamics in play is that of a system that is (tunably) ``small'' compared to the full (``large $N$'') number of degrees of freedom usually described by these gravity duals. Putting this together, as done here, holographic heat engines that employ this reduced system would seem to define models of quantum heat engines. We showed that the black hole Otto cycle, presented here for the first time, has exceedingly interesting properties in this regard. This connection may be of use because it enriches the family of quantum heat engines that have been defined so far, and ---perhaps more importantly--- provides an entirely new laboratory of techniques for exploring the thermodynamics of small quantum systems. 

It would be interesting to explore this connection further, and there are many fruitful avenues along which to depart. Besides constructing more examples (there is a host of black hole solutions of various types, in various dimensions, that have $C_V\neq0$) it would be interesting to directly employ the underlying quantum description ({\it i.e.,} holography) to explore features of the engines. An effective field theory model of the reduced system that operates at fixed thermodynamic volume would be useful in this regard, and it can presumably be derived directly from the holographically dual field theory. (It may begin with a system of, possibly truncated,  quantum harmonic oscillators in some number of dimensions, given the observations in section~\ref{sec:schottky-peaks}. Perhaps the thermodynamic volume their inverse  effective frequency, generalizing the 1--dimensional case reviewed in section~\ref{sec:quantum-heat-engines}.) The thermalization process that happens at constant volume may be illuminated by studying the evolution of the entanglement entropy. This is again something that is accessible in the holographic description, using the techniques of refs.\cite{Ryu:2006bv,Ryu:2006ef}, suitably adapted to constant volume processes. As mentioned in section~\ref{sec:synthesis}, getting at the underlying quantum features of the system in this way may be useful for addressing the issue of how quantum heat engines differ from their classical counterparts in certain regimes, such as considering operation at finite power.

The quantum heat engine models that are accessible using our construction are in general  likely to be somewhat more complex  (or simple, depending upon point of view) than those that have been extensively studied in the literature so far. (It would be an unexpected, though pleasant, surprise to be able to model a pure two--level spin system with a black hole, for example.) As discussed in section~\ref{sec:kerr-otto}, this is partly due to the fact that we have selected a small number of degrees of freedom out a very large parent system, resulting in very large $\gamma=C_p/C_V$. It would be of interest to try to see if there are gravitational models with more parameters that can help avoid this limitation, giving access to the small $\gamma$s (of order 1) of familiar materials while retaining the small system limit. This does not mean that the construction may not be a useful laboratory for testing ideas and even modelling phenomena seen in experiments. 

Indeed, since quantum heat engines are of considerable active experimental interest, with new realizations appearing regularly in a range of contexts (spin systems, trapped ions, resonant cavities, {\it etc.}), the working substances being used will become increasingly more intricate, and may well yield exotic behaviours. A holographic heat engine description {\it via} black holes may well prove useful for understanding their features, perhaps in ways analogous to how large $N$ models gave useful insights into strongly coupled aspects of nuclear physics, and condensed matter physics.

 
\section*{Acknowledgements}
The work of CVJ  was funded by the US Department of Energy  under grant DE-SC 0011687.  CVJ would  like to thank  Lincoln Carr and the other organizers of the KITP conference ``Open Quantum Systems in Quantum Simulators'' for the invitation to present some of this work\cite{conference}, and   many  participants for their warm welcome and interest.  Thanks, as always, to Amelia for her support and patience.

\bibliographystyle{utphys}
\bibliography{black_hole_quantum_heat_engines}

\end{document}